\documentclass[twocolumn, groupedaddress, amsmath, amssymb, aps, prl, floatfix]{revtex4-2}
\usepackage{graphicx}% Include figure files
\usepackage{dcolumn}% Align table columns on decimal point
\usepackage{bm}% bold math
\usepackage{verbatim}% comment environment
\usepackage{hyperref}% add hypertext capabilities
\usepackage{color}
\usepackage{enumitem}

\usepackage[resetlabels,labeled]{multibib}
\newcites{S}{supplement references}

\begin{document}

\title{Rotor Lattice Model of Ferroelectric Large Polarons}
\author{Georgios M. Koutentakis}
\affiliation{Institute of Science and Technology Austria (ISTA), am Campus 1, 3400 Klosterneuburg, Austria}
\author{Areg Ghazaryan}
\affiliation{Institute of Science and Technology Austria (ISTA), am Campus 1, 3400 Klosterneuburg, Austria}
\author{Mikhail Lemeshko}
\affiliation{Institute of Science and Technology Austria (ISTA), am Campus 1, 3400 Klosterneuburg, Austria}
\date{\today}

\begin{abstract}
We present a minimal model of charge transport in hybrid  perovskites, which provides an intuitive explanation for the recently proposed formation of ferroelectric large polarons. We demonstrate that short-ranged charge--rotor interactions lead to long-range ferroelectic ordering of rotors, which strongly affects the carrier mobility. In the nonperturbative regime, where our theory  cannot be reduced to any of the earlier models,  we predict polaron properties in good agreement with experiment. This shows the potential of simple models to reveal electronic properties of molecular materials.
\end{abstract}

\maketitle

Hybrid organic-inorganic perovskites (HOIP) are praised for their outstanding performance in photovoltaic applications  due to their long carrier lifetimes and diffusion lengths~\cite{brenner2016hybrid, johnston2016hybrid,jena2019halide}. After the initial reports on spectacular optoelectronic properties of HOIP solar cells~\cite{kojima2009organometal,chung2012all,lee2012efficient,kim2012lead}, the field has expanded at an unprecedented pace~\cite{miyasaka2021perovskite,fujiwara2022hybrid}.  It turned out that the physical properties of HOIP are quite complex, \textit{inter alia} due to their soft structure~\cite{miyata2017lead}, ionic mobility~\cite{eames2015ionic,yuan2016ion}, and the interplay between rotational dynamics of the molecular cations and their structural and  (photo)electric  properties~\cite{chen2017origin, selig2017organic,liu2022effects,mozur2021cation}. Despite their complexity, a considerable progress in understanding of HOIP has been achieved through density functional theory~\cite{even2014dft,even2015solid,yun2017theoretical,traore2022band}, molecular dynamics~\cite{meggiolaro2020polarons} and machine learning~\cite{jinnouchi2019phase,zhang2020machine,myung2022challenges} approaches. Based on such atomistic simulations it is, however, challenging to obtain a simple intuitive picture independent of microscopic details,  motivating the development of minimal models that capture the key physical properties of HOIP.

Perhaps the most pressing issue to be addressed in HOIP concerns their charge transport properties. Although the carrier recombination lifetimes and diffusion lengths in HOIP are  comparable to that of conventional semiconductors such as GaAs, the charge mobilities are orders of magnitude smaller~\cite{brenner2015mobilities}.  Earlier theories tried to explain this  through screening of excitons by collective orientation of organic cations~\cite{even2014analysis} and modification of the band edges due to spin-orbit coupling resulting from cation-induced structural variability~\cite{amat2014cation,even2014dft,zheng2015rashba}. In contrast, recent theories  emphasize the role of the large polarons   screening the carriers from charged defects, other carriers, and phonons~\cite{zhu2015charge,welch2016density,neukirch2016polaron,ivanovska2017long,zheng2019large,ambrosio2018origin,ambrosio2019charge,wang2022phonon, miyata2017large,miyata2018ferroelectric,wang2020solvated}.  Still, the detailed origin of large polaron formation remains debated.  A promising approach \cite{miyata2018ferroelectric,wang2020solvated} suggests ferroelectric large polarons, consisting of ferroelectrically ordered nanodomains, which are postulated to provide substantially increased screening compared to Fr{\" o}hlich polarons~\cite{Froehlich1954,AlexandrovDevreese2010}.  However, the involved properties of HOIP  make quantitative predictions for such quasiparticles difficult~\cite{miyata2018ferroelectric}, hindering their unambiguous experimental identification.

\begin{figure}[t]
\centering
\includegraphics[width=1\linewidth]{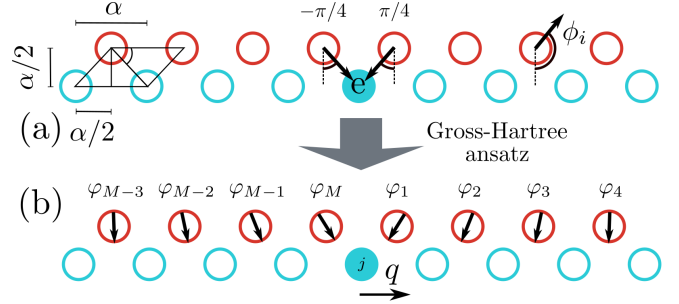}
\caption{(a) The  tight binding model. Blue empty (filled) circles label  empty (occupied) electron sites. Red circles show the dipole positions modeled by planar rotors. The distances between the sites and the orientations of the rotors/dipoles are also shown,  where $\alpha$ is the lattice constant. (b) Illustration of the Gross-Hartree ansatz. The electron possesses quasimomentum, \(q\), while the rotor states relative to it are described by the single-rotor states, \(\varphi_j(\phi)\). Arrows show the orientations of the rotors. \label{fig_schematic}}
\end{figure}

In this Letter we show that the formation of ferroelectric large polarons  takes place already in a minimal model, where charge carriers interact with an one-dimensional array of planar rotors, see Fig.~\ref{fig_schematic}(a).  The dipolar rotors model the reorientation dynamics of  organic molecular cations, $\mathrm{A}^+$ in the $\mathrm{ABX}_3$ perovskite structure.  Molecules interact with charge carriers hopping on the inorganic sublattice made of octahedral $\mathrm{BX}_6^{-}$ cages, which we represent by discrete sites. Due to screening \cite{ZhuMiyata2016}, we assume  charge--dipole interactions to be short-ranged  and dipole--dipole interactions to be absent.   As we demonstrate, this model captures the formation of local  $\sim$10~nm-sized ferroelectric order \cite{RakitaOmri2017,ShahrokhiGao2020,miyata2017large} and the crossover between a large light polaron (associated with ferroelectrically polarized dipoles) and a small heavy polaron regime characterized by charge carrier localization at the boundary of two misaligned ferroelectrically ordered domains \cite{LiuCollins2018,wang2020solvated}.
 
This ferroelectric order significantly increases the effective mass of the carriers even within the light polaron regime,  in agreement with the modest but not negligible mobilities observed in HOIP \cite{brenner2015mobilities},  but in contrast to other polaron models, e.g. the Holstein polaron, which predict much larger renormalization \cite{AlexandrovYavidov2004, TozerBarfold2014}. When the domain wall forms, the effective mass grows exponentially, suggesting high anisotropy of the mobilities and diffusion constants of large ferroelectric polarons in two and three dimensions. Anisotropies along different crystalographic directions have been recently experimentally identified~\cite{TailorSatapathi2022,BaimuratovPereziabova2017,JiaoYi2021}, confirming the relevance of our model for HOIP.

The Hamiltonian of our model, cf.\ Fig.~\eqref{fig_schematic}(a), reads
\begin{multline}
        \hat{H} = -t \sum_{i=1}^{M} (\hat{a}_{i+1}^{\dagger} \hat{a}_i + {\rm h.c.})
            - B \sum_{i = 1}^{M} \frac{\partial^2}{\partial \phi_i^2} \\
            - V_0 \sum_{i=1}^{M} \hat{a}_{i}^{\dagger} \hat{a}_i \left[ \cos\left( \phi_i + \frac{\pi}{4} \right)
            +\cos\left( \phi_{i - 1} - \frac{\pi}{4} \right) \right],
\label{Hamilt}
\end{multline}
where \(\hat{a}_i\) (\(\hat{a}_i^{\dagger}\)) are the electron annihilation (creation) operators, angles \(\phi_i\) define the dipole orientations and $B$ their rotational constants (in what follows we  use the terms dipoles and rotors interchangeably), \(t\) is the tunneling rate of the electron,   \(V_{0}\) is the electron--dipole interaction strength and \( M \) is the number of  rotors in the lattice. For simplicity we neglect the activation energy of molecular rotations, $E_{\rm act}$, as its presence effectively inhibits rotor--electron interactions for $E_{\rm act} \gtrsim V_0$~\cite{FabiniSiaw2017,liu2022effects}. Note that  we employ periodic boundary conditions, i.e. \( \hat{a}_{M+j}^{\dagger} = \hat{a}_{j}^{\dagger} \) and \( \phi_{M+j} = \phi_j \), for all $j = 1, \dots, M$. Although the model can be trivially extended to hole carriers by assuming $V_0 < 0$, here we focus on electrons,  $V_0 > 0$.

In HOIP the molecular rotational energy  $B \sim 1~{\rm meV}$ is the lowest energy scale  since $V_0, t \sim 0.1-1~{\rm eV}$ \cite{Kang2017dynamic, FabiniSiaw2017}. To generate an appropriate rotor basis for \(B \ll t\), we variationally optimize the state of the rotors relative to the electron, $\varphi_j(\phi)$, cf.\ Fig. \ref{fig_schematic}(b), based on the following ansatz:
\begin{equation}
\begin{split}
| \Psi_q (\phi_1, \dots, \phi_M) \rangle &=\sum_{j = 1}^M \frac{e^{i \frac{2 \pi q}{M} j}}{\sqrt{M}}  \prod_{k = 1}^M \varphi_{I(k,j)} (\phi_k) \hat{a}^{\dagger}_j | 0 \rangle,
\end{split}
\label{GH-variational-ansatz}
\end{equation}
where \(| 0 \rangle\) and \(\hat{a}_j\) are the electron vacuum and creation operators and \(q = 0,\dots,M-1 \) gives the  quasimomentum of the polaron state. We will  refer to this approach as the variational Gross-Hartree method (vGH). The indices of \(\varphi_j(\phi)\) appearing in Eq.~\eqref{GH-variational-ansatz} read \( I(k,j) = 1 + [(M + k - j) \bmod M] \) and are selected such that the rotor state depends only on the relative distance between the rotor and the electron. For instance, \(\varphi_1(\phi)\) and \(\varphi_M(\phi)\) refer to the state of the rotor on the right and left of the electron, respectively, independently of the position of the latter, cf.\ Fig. \ref{fig_schematic}(b).

Note that while the ansatz of Eq.~\eqref{GH-variational-ansatz} generalizes the basis generated via the Lang-Firsov transformation \cite{AlexandrovDevreese2010,LangFirsov1963} as it allows for $t$-dependent modifications of the rotor state, it neglects dipole--dipole correlations. These are not expected to limit the applicability of vGH, since no direct interaction between dipoles appears in Eq.~\eqref{Hamilt}. Thus, only dipole--dipole correlations mediated by the electron can take place, which result in small corrections  in related polaron models (except for strong coupling)~\cite{grusdt2015new}. The applicability of the vGH approximation has been justified through comparison with exact diagonalization for small $M$~\footnote{See Supplemental Material for details.}.

\begin{figure}[t]
\centering
\includegraphics[width=1\linewidth]{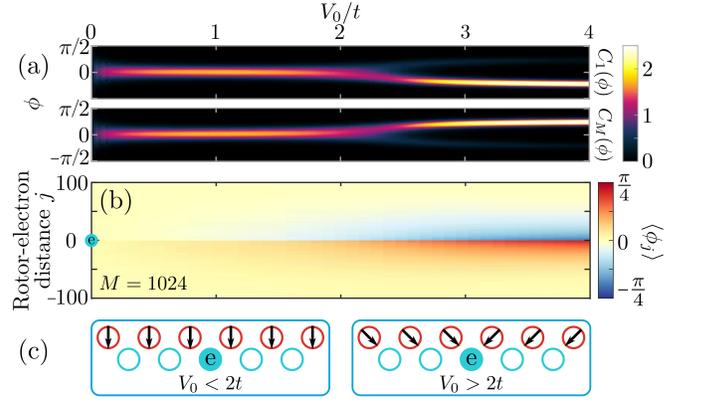}
\caption{(a)  Rotor-electron correlation function, $C_j(\phi)$, for $j = 1$ and $j=M=4$, as a function of the electron--rotor interaction, $V_0$. The electron is fixed at the first site.
(b)~Average rotor orientation, $\langle \phi_j \rangle$, depending on the electron--rotor distance and $V_0$ for  \( M = 1024 \).
In both cases $B = 10^{-3}t$.
(c)~Schematic illustration of the   ferroelectric orders involved.
\label{fig_polaron_density}}
\end{figure}

The  order emerging in the rotor lattice  can be elucidated by considering the  rotor--electron correlation function at distance $j$, $C_j(\phi) = | \varphi_j(\phi) |^2$, as shown in Fig.~\ref{fig_polaron_density}(a) for a small system with $M = 4.$ In Fig.~\ref{fig_polaron_density}(b) we provide the average polarization of the rotors, $\langle \phi_j \rangle = \int_{-\pi}^{\pi} \mathrm{d}\phi~\phi C_j(\phi)$ for  $M=1024$, which is large enough to achieve convergence towards the $M \to \infty$ limit.
 
In the case of small \( B = 10^{-3} t \), relevant for HOIP, we observe the emergence of two distinct interaction regimes. For \( V_0 < 2 t \) the rotors become strongly polarized towards the electronic lattice, \( \phi \approx 0 \), and an almost perfect ferroelectric order emerges (see Fig.~\ref{fig_polaron_density}(a) and the left panel of Fig.~\ref{fig_polaron_density}(c)). Similarly, Fig.~\ref{fig_polaron_density}(b) shows $\langle \phi_j \rangle \approx 0$, for all $j$ within this $V_0$ range. For \( V_0  > 2 t \), on the other hand, we observe  domain formation in the rotor system, see Fig.~\ref{fig_polaron_density}(a) and the right panel of Fig.~\ref{fig_polaron_density}(c). The rotors to the left of the electron, \( M/2 < j \le M \), polarize with \( \phi \approx \pi /4 \), while the rotors at \(1 \le j \le M/2 \)  polarize towards \( \phi \approx - \pi/4 \).  Thus, the electron acts as a ferroelectric domain wall, with rotors on each side of the electron pointing towards it. From Fig.~\ref{fig_polaron_density}(b) we can see that although this rotor ordering is local, it is quite extensive involving $\sim 50$ rotors in each side of the electron. Note that this change in ferroelectric order with varying $V_0$ is gradual, of typical crossover character \cite{Spohn1986, GerlachLowen1991},  as the rotors neighbouring the electron from either side possess slightly different average orientations even for $V_0 < 2 t$, see Fig.~\ref{fig_polaron_density}(b).

The interaction dependence of the local ferroelectric order, Fig.~\ref{fig_polaron_density}, provides an intuitive picture for the role of molecular dipole moments in the formation of polarons at ferroelectric domain boundaries, proposed in Ref.~\cite{wang2020solvated}.  That work suggests that the carriers are confined to and move along a two-dimensional ferroelectric domain-wall, whereas the   hopping perpendicular to it is much slower due to the distortion of the inorganic lattice. In our model this distortion corresponds to a reduced $t$ along the distortion direction, resulting in an effectively higher $V_0/t$ that can exceed the threshold for formation of ferroelectric domain walls, $V_0 \approx 2 t$ in one dimension. In contrast, along the directions where no  distortion takes place, $V_0/t$ remains smaller than the threshold, which stabilizes an almost perfectly polarized rotor state.

\begin{figure}[t]
\centering
\includegraphics[width=1\linewidth]{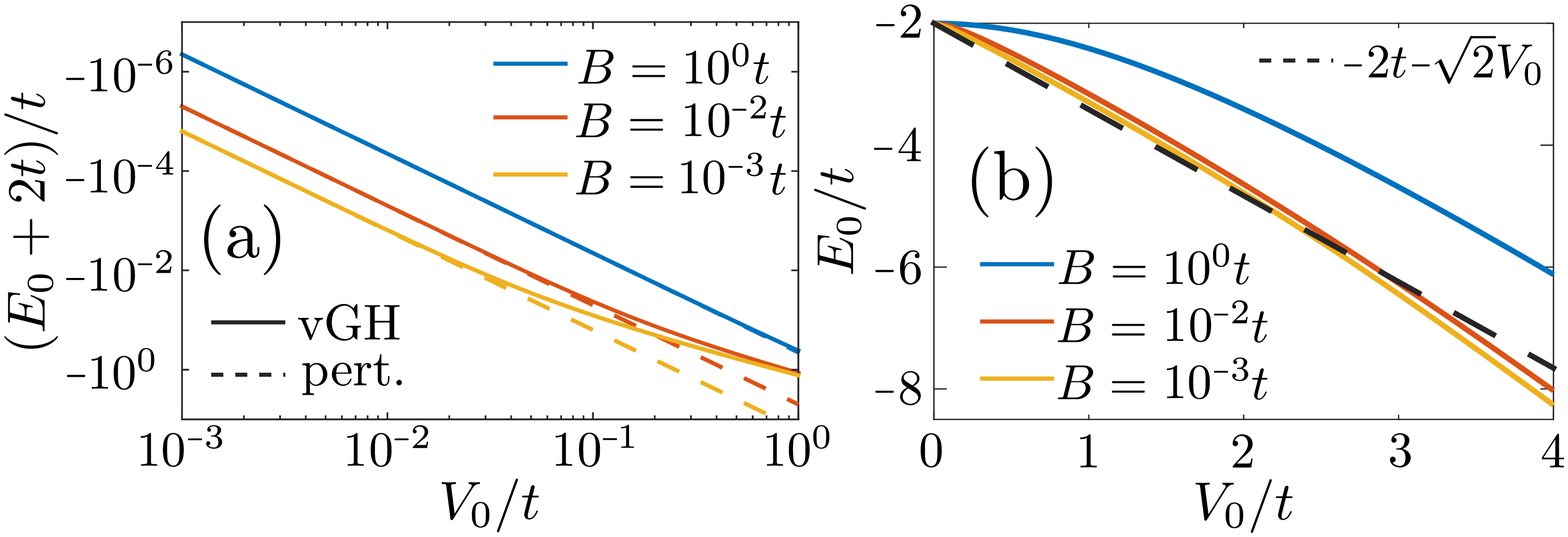}
\caption{
(a, b) Polaron energy, $E_0$, for different values of $B$ as a function of   $V_0$. (a) compares the vGH results (solid lines)  with perturbation theory (dashed lines). The dashed line in (b) is an eye-guide to estimate $E_0$ (see the text). In all cases \(M = 1024\).
\label{fig_polaron_energy}}
\end{figure}

The origin of the emerging order can be elucidated by examining the polaron energy, $E_0$. First, let us analyze its scaling with $V_0$,  Fig.~\ref{fig_polaron_energy}(a). For small $V_0$,  the polaron energy follows the pertubative result, $E_0 = -2 t -V_0^2/\sqrt{B(B+4t)}$,  independently of $B$~\cite{Note1}. With increasing $V_0$, however, the energy of the polaron diverges from this scaling, with the strongest deviations observed for smaller $B$'s. This behavior stems from the breakdown of perturbation theory for $V_0 > 2 \sqrt{B t}$, where the rotor--electron interaction creates a large number of rotor excitations.

The fact that the ferroelectric dressing of the electron observed in Fig.~\ref{fig_polaron_density} takes place beyond the regime of validity of perturbation theory, implies that it originates from the collective excitations of the rotor array and their coupling to the electron.  Since the spectrum of rotors is different from that of harmonic oscillators, the nonperturbative physics of the ferroelectric polaron given by Eq.~\eqref{Hamilt} is fundamentally different from the traditional  models such as the Holstein polaron \cite{AlexandrovDevreese2010, Holstein1959a, Holstein1959b}.

The fundamental difference between the $B < 10^{-2} t$ and $B \approx t$ regimes is directly observable by comparing the polaron energies for different $B$'s  near  the crossover point, $V_0 \sim 2t$, see Fig.~\ref{fig_polaron_energy}(b). For  \( B < 10^{-2} t \) and $V_0 < 2 t$, the polaron energy   features an almost linear decrease, $E_0 \approx -2 t - \sqrt{2} V_0$, stemming from  strong polarization of the rotors ($\phi \approx 0$) in the vicinity of the electron. This results in the potential energy contribution $\sim V_0 [\cos(\pi/4) + \cos(-\pi/4)] = - \sqrt{2} V_0$.  For stronger interactions, $V_0 > 2 t$, the polaron energy decreases faster than $\propto -\sqrt{2} V_0$ due to  the domain-wall formation at the electron positions, which  increases the rotor--electron attraction. For  $B \sim t$, the behaviour of the system changes and the polaron energy decreases quadratically.  This is due to the  large amount of energy required to create rotor excitations which hinders their polarization and the associated potential energy benefit, thereby precluding the formation of ferroelectric order.

\begin{figure*}[t]
\centering
\includegraphics[width=1\linewidth]{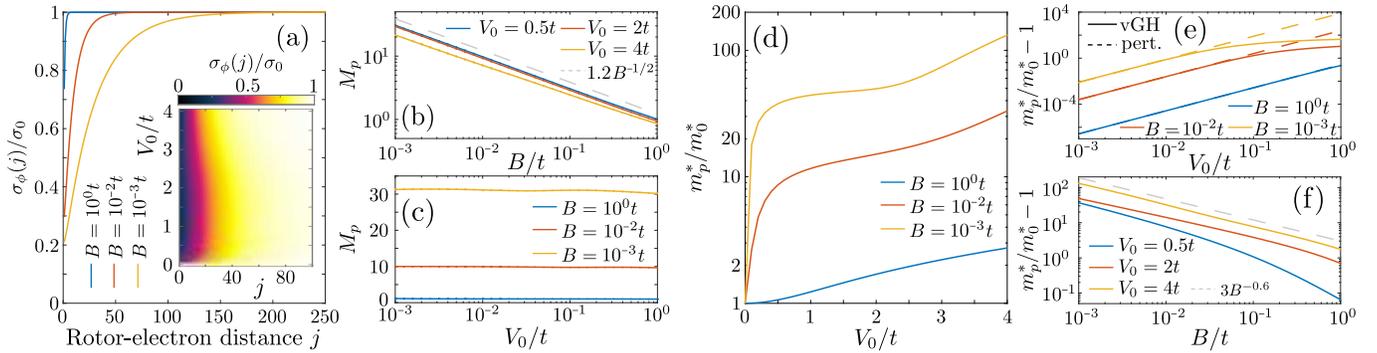}
\caption{
(a) Variance of the rotor orientations, $\sigma_\phi(j)/\sigma_0$, for different \( B \)  and \( V_0/t = 0.5 \). Inset: \( V_0 \) dependence of $\sigma_\phi(j)/\sigma_0$ for \( B= 10^{-3} t \). (b, c) Polaron size, $M_p$, derived from exponential fits of $\sigma_\phi(j)/\sigma_0$ as a function of (b) $B$ and (c) $V_0$. (d, e) The ratio of the polaron and free electron effective masses, \( m^{*}_{p}/m^*_0 \) for different values of \( B \), as a function of \(V_0 \). (e) Comparison of vGH results (solid lines) to  perturbation theory (dashed lines). (f) The ratio \( m^{*}_{p}/m^*_0 \) as a function of $B$. The dashed line in (b) and (f) serves as an eye-guide to estimate  the scaling  with $B$.  In all cases \( M = 1024 \).
\label{fig_effective_mass}}
\end{figure*}

Having discussed the basic properties of the polaronic states  based on energetic arguments, let us  focus on their coherence and transport properties and their relation to HOIP experiments. The polaron size, or, equivalently, the polaron coherence length, is associated with the extent of the ferroelectric order in the vicinity of the electron, see also Fig.~\ref{fig_polaron_density}(b). This property can be expressed through the variance of the rotor angle, \( \sigma_{\phi}(j) = \left[\int_{-\pi}^{\pi} \mathrm{d}\phi~(\phi - \langle \phi_j \rangle)^2 C_j(\phi)\right]^{1/2} \), over the variance of the uniform distribution, \( \sigma_{0} =  2 \pi/\sqrt{12}\). The  value of $\sigma_{\phi}(j)/\sigma_0 = 1$ corresponds to a uniform density profile, where the   dipoles are unaffected by the electron motion. The values \( 0 \leq \sigma_{\phi}(j)/\sigma_0 <1\) correspond to the polarization of the $j$-th rotor typical for ferroelectric order.

The localization of ferroelectric order is demonstrated by the exponential trend of $\sigma_\phi(j)/\sigma_0$ which rapidly saturates to unity as rotors far away from the electron remain not oriented, see Fig. \ref{fig_effective_mass}(a). For fixed $V_0$, a larger number of rotors can be excited  at smaller \( B \), giving rise to a more extensive dressing cloud around the electron. Inversely, for \( t = B \), the polaron  is  strongly localized in the vicinity of the  electron, \( j \to 0 \), and thus cannot be called a large polaron. This is fully consistent with the  energetic arguments presented above, cf.\ Fig.~\ref{fig_polaron_energy}(b), and is further illustrated in Fig.~\ref{fig_effective_mass}(b). Here we use an exponential fit, $\sigma_{\phi}(j)/\sigma_0 = 1 + (\sigma_{\phi}(1)/\sigma_0 -1) e^{-(j-1)/M_p}$ for $j < M/2$, to extract the polaron size $M_p$. We find that almost independently of $V_0$, the polaron size grows as $\propto B^{-1/2}$ with decreasing~$B$.

In contrast, for a constant $B$, the spatial extent of the polaron depends weakly on $V_0$, see the inset of Fig. \ref{fig_effective_mass}(a), especially within the polarized regime, \( V_0 < 2t \). This observation is confirmed by Fig. \ref{fig_effective_mass}(c), which shows that $M_p$ is independent of $V_0$ even for very small $V_0$'s. This can be explained along the lines of  perturbation theory: electron--rotor interactions result in virtual rotor excitations  localized in the vicinity of the electron, whose momentum shifts from $q$ to $q'$. These excitations are characterized by an energy   $B$ and thus a lifetime $\sim \hbar/B$. Consequently, the maximum distance between the electron and an excitation depends solely on the distribution of available $q'$ and the excitation lifetime, both of which are independent of $V_0$ controlling the excitation probability. For  $V_0 > 2 t$, where the ferroelectric domain wall forms, the spatial extent of the polaron decreases by a factor of $\sim$2, see the inset of Fig.~\ref{fig_effective_mass}(a). In this regime, the perturbative argumentation is invalid, since, as argued below, the electron becomes rigidly attached to its dressing cloud of rotor excitations.  In summary, although the ferroelectric dressing is found to be large at the level of the unit cell, its spatial extent, $M_p \times \alpha \approx 30 \times 5 \text{\r{A}} \approx 15$~nm, is much smaller than the observed diffusion lengths of $> 1$~$\mu$m \cite{StranksEperon2013,ShiAdinolfi2015,DongFang2015}. This implies that the semiclassical treatment of diffusion lengths, frequently used in the literature \cite{brenner2015mobilities}, is well justified within our model.

The ferroelectric order crucially affects the polaron mass, Fig.~\ref{fig_effective_mass}(d). Note that the  mass scale for $V_0 = 0$ is $m^*_0 = \hbar^2/(2 t \alpha^2) \approx 0.15~m_e$ (for $\alpha \approx 5~\text{\r{A}}$ and $t \approx 1~\text{eV}$ relevant for HOIP).  For smaller \( B \),  the effective mass features a strong overall increase.  For $B < 10^{-2} t$ the initial growth  of $m_p^* = (\hbar^2/\alpha^2) (\partial^2 E_p/\partial q^2)^{-1}$ at smaller \( V_0 \) is followed by a plateau at  \(V_0 \sim t\).  This can be rationalized  by considering how $m_p^*$, scales with $B$ and $V_0$. From Fig.~\ref{fig_effective_mass}(e)  we see that for small $V_0$ the effective mass increases following  the perturbative result, $m_p^*/m_0^* = 1 + V_0^2 (B + 2 t)/[B (B + 4 t)]^{3/2}$~\cite{Note1}, and saturates at larger $V_0$. The saturation of $m_p^*$ can be thought of as an almost rigid attachment of the ferroelectric polarization cloud to the electron at strong interactions.  For $B \sim t$ the attachment is precluded by  rapid rotation of the rotors, resulting in no saturation of $m_p^*$. The scaling of $m_p^*$ with $B$, Fig.~\ref{fig_effective_mass}(f), demonstrates significant deviations from the perturbative result, $m_p^*/m_0^* - 1 \propto B^{-3/2}$, in the region where the ferroelectric polaron forms, $V_0 \sim t \gg B$. The $B$-scaling is found to be significantly less steep, $m_p^*/m_0^* - 1 \propto B^{-0.6}$ for $V_0 = 0.5 t, 4 t$, and $m_p^*/m_0^* - 1 \propto B^{-0.5}$ in the crossover region $V_0 = 2t$. Importantly, $m_p^*$ is a decreasing function of $B$ in all of the considered cases.

Thus, although the polaron for $V_0 < 2 t$  is large, it features low but non-negligible mobility, $\mu \propto \tau/m_p^*$, which is consistent with HOIP experiments~\cite{brenner2015mobilities}. Here, $\tau$ corresponds to the mean scattering time which is large in HOIP \cite{Price2015} and is expected to increase due to polaron screening.  Also, we would like to emphasize that the simultaneous power-law increase of the polaron size, $M_p$, and of its effective mass, $m_p^*$, with decreasing $B$, see Fig.~\ref{fig_effective_mass}(b, f), is the behaviour that sets our model apart from the well-known Holstein and Fr{\" o}hlich polarons, where larger effective masses are associated with smaller polaron sizes \cite{AlexandrovDevreese2010, Holstein1959a, Holstein1959b, JackelmannWhite1998} or extremely heavy polarons with negligible mobilities~\cite{AlexandrovYavidov2004, TozerBarfold2014}. This indicates that the framework of rotor lattices introduced here has the potential to  explain the apparently contradicting features of carrier dynamics in HOIP.

For  \( V_0 > 2t \), $m^*_p$ grows exponentially with \( V_0 \), Fig.~\ref{fig_effective_mass}(d), as a consequence of the reduced mobility due to the domain wall  co-moving with the electron. This is consistent with strong anisotropy of the effective mass along vs.\ perpendicular to the ferroelectric domain wall once it forms~\cite{TailorSatapathi2022,BaimuratovPereziabova2017,JiaoYi2021}. Our results further suggest that the anisotropy in coherence length is much less pronounced, see the inset of Fig.~\ref{fig_effective_mass}(a). Thus the study of the relation between the coherence and mobility anisotropies might be important for the experimental detections of ferroelectric polarons.

In conclusion, we  proposed a minimal, tractable, and extendable rotor lattice model describing the formation of ferroelectric polarons in HOIP \cite{miyata2017large,miyata2018ferroelectric,wang2020solvated}. The model captures several observed features of polarons  such as their modest mobility but large coherence length. Furthermore, it provides intuition for the mechanism behind large polaron formation at ferroelectric domain boundaries, proposed in Ref.~\cite{wang2020solvated}.  Our model lays the groundwork for realizing a top-down approach to the carrier dynamics in HOIP, complementary to the existing density functional theory studies \cite{even2014dft,even2015solid,yun2017theoretical,traore2022band}. Possible extensions include studying  the phase diagram of the two-dimensional lattice system where signatures of carrier localization along different directions can be identified.  The study of electron--hole interactions mediated by the rotors can elucidate the impact of the molecules on the observed long carrier lifetimes. Moreover, our study suggests an interesting interplay of mobility inhomogeneity and exciton lifetime which might provide quantitative predictions for the diffusion length. Different forms of rotor--phonon coupling Hamiltonians can be accounted for in order to fully capture the ferroelectric properties of HOIP materials. In addition, studies that connect the abstract model parameters with realistic material properties will be crucial. The possibility of angulon formation   affecting  molecular mobility~\cite{schmidt2015rotation,schmidt2016deformation, YakaboyluScrPRL17, cui2022microscopic,wu2022optical} and of  the halogen-metal hybridization which can introduce polarization of the ${\rm BX}_6^-$ cages \cite{Volosniev22a, Volosniev22b} might also be relevant for reliable modeling of HOIP properties with rotor lattice setups.

\begin{acknowledgments}
We thank Zhanybek Alpichshev, Artem Volosniev and Alexandra V. Zampetaki for fruitful discussions and comments. This project  received funding from the European Union’s Horizon 2020 research and innovation programme under the Marie Skłodowska-Curie grant agreement No.~ 101034413. M.L.~acknowledges support by the European Research Council (ERC) Starting Grant No.~801770 (ANGULON).
\end{acknowledgments}

\bibliographystyle{apsrev4-1}
\bibliography{bibliography}

\begin{thebibliography}{19}%
\makeatletter
\providecommand \@ifxundefined [1]{%
 \@ifx{#1\undefined}
}%
\providecommand \@ifnum [1]{%
 \ifnum #1\expandafter \@firstoftwo
 \else \expandafter \@secondoftwo
 \fi
}%
\providecommand \@ifx [1]{%
 \ifx #1\expandafter \@firstoftwo
 \else \expandafter \@secondoftwo
 \fi
}%
\providecommand \natexlab [1]{#1}%
\providecommand \enquote  [1]{``#1''}%
\providecommand \bibnamefont  [1]{#1}%
\providecommand \bibfnamefont [1]{#1}%
\providecommand \citenamefont [1]{#1}%
\providecommand \href@noop [0]{\@secondoftwo}%
\providecommand \href [0]{\begingroup \@sanitize@url \@href}%
\providecommand \@href[1]{\@@startlink{#1}\@@href}%
\providecommand \@@href[1]{\endgroup#1\@@endlink}%
\providecommand \@sanitize@url [0]{\catcode `\\12\catcode `\$12\catcode
  `\&12\catcode `\#12\catcode `\^12\catcode `\_12\catcode `\%12\relax}%
\providecommand \@@startlink[1]{}%
\providecommand \@@endlink[0]{}%
\providecommand \url  [0]{\begingroup\@sanitize@url \@url }%
\providecommand \@url [1]{\endgroup\@href {#1}{\urlprefix }}%
\providecommand \urlprefix  [0]{URL }%
\providecommand \Eprint [0]{\href }%
\providecommand \doibase [0]{http://dx.doi.org/}%
\providecommand \selectlanguage [0]{\@gobble}%
\providecommand \bibinfo  [0]{\@secondoftwo}%
\providecommand \bibfield  [0]{\@secondoftwo}%
\providecommand \translation [1]{[#1]}%
\providecommand \BibitemOpen [0]{}%
\providecommand \bibitemStop [0]{}%
\providecommand \bibitemNoStop [0]{.\EOS\space}%
\providecommand \EOS [0]{\spacefactor3000\relax}%
\providecommand \BibitemShut  [1]{\csname bibitem#1\endcsname}%
\let\auto@bib@innerbib\@empty
%</preamble>
\bibitem [{\citenamefont {Alexandrov}\ and\ \citenamefont
  {Devreese}(2010)}]{AlexandrovDevreese2010s}%
  \BibitemOpen
  \bibfield  {author} {\bibinfo {author} {\bibfnamefont {A.~S.}\ \bibnamefont
  {Alexandrov}}\ and\ \bibinfo {author} {\bibfnamefont {J.~T.}\ \bibnamefont
  {Devreese}},\ }\href {\doibase 10.1007/978-3-642-01896-1_1} {\emph {\bibinfo
  {title} {Advances in Polaron Physics}}}\ (\bibinfo  {publisher} {Springer},\
  \bibinfo {address} {Berlin Heidelberg,~},\ \bibinfo {year}
  {2010})\BibitemShut {NoStop}%
\bibitem [{\citenamefont {Lang}\ and\ \citenamefont
  {Firsov}(1963)}]{LangFirsov1963s}%
  \BibitemOpen
  \bibfield  {author} {\bibinfo {author} {\bibfnamefont {I.~G.}\ \bibnamefont
  {Lang}}\ and\ \bibinfo {author} {\bibfnamefont {Y.~A.}\ \bibnamefont
  {Firsov}},\ }\href
  {http://www.jetp.ras.ru/cgi-bin/e/index/r/45/1/p378?a=list} {\bibfield
  {journal} {\bibinfo  {journal} {Zh. Eksp. Teor. Fiz.}\ }\textbf {\bibinfo
  {volume} {45}},\ \bibinfo {pages} {378} (\bibinfo {year} {1963})},\ \bibinfo
  {note} {[Sov. Phys. JETP 18, 262 (1964)]}\BibitemShut {NoStop}%
\bibitem [{\citenamefont {Holstein}(1959{\natexlab{a}})}]{Holstein1959as}%
  \BibitemOpen
  \bibfield  {author} {\bibinfo {author} {\bibfnamefont {T.}~\bibnamefont
  {Holstein}},\ }\href {\doibase https://doi.org/10.1016/0003-4916(59)90002-8}
  {\bibfield  {journal} {\bibinfo  {journal} {Ann. Phys.}\ }\textbf {\bibinfo
  {volume} {8}},\ \bibinfo {pages} {325 } (\bibinfo {year}
  {1959}{\natexlab{a}})}\BibitemShut {NoStop}%
\bibitem [{\citenamefont {Holstein}(1959{\natexlab{b}})}]{Holstein1959bs}%
  \BibitemOpen
  \bibfield  {author} {\bibinfo {author} {\bibfnamefont {T.}~\bibnamefont
  {Holstein}},\ }\href {\doibase https://doi.org/10.1016/0003-4916(59)90003-X}
  {\bibfield  {journal} {\bibinfo  {journal} {Ann. Phys.}\ }\textbf {\bibinfo
  {volume} {8}},\ \bibinfo {pages} {343 } (\bibinfo {year}
  {1959}{\natexlab{b}})}\BibitemShut {NoStop}%
\bibitem [{\citenamefont {Abramowitz}\ and\ \citenamefont
  {Stegun}(1965)}]{AbramowitzStegun}%
  \BibitemOpen
  \bibinfo {editor} {\bibfnamefont {M.}~\bibnamefont {Abramowitz}}\ and\
  \bibinfo {editor} {\bibfnamefont {I.~A.}\ \bibnamefont {Stegun}},\ eds.,\
  \href@noop {} {\emph {\bibinfo {title} {Handbook of mathematical
  functions}}},\ Dover Books on Mathematics\ (\bibinfo  {publisher} {Dover
  Publications},\ \bibinfo {address} {Mineola, NY},\ \bibinfo {year}
  {1965})\BibitemShut {NoStop}%
\bibitem [{\citenamefont {Huba{\v{c}}}\ and\ \citenamefont
  {Wilson}(2010)}]{Hubac2010}%
  \BibitemOpen
  \bibfield  {author} {\bibinfo {author} {\bibfnamefont {I.}~\bibnamefont
  {Huba{\v{c}}}}\ and\ \bibinfo {author} {\bibfnamefont {S.}~\bibnamefont
  {Wilson}},\ }\enquote {\bibinfo {title} {Brillouin-wigner perturbation
  theory},}\ in\ \href {\doibase 10.1007/978-90-481-3373-4_2} {\emph {\bibinfo
  {booktitle} {Brillouin-Wigner Methods for Many-Body Systems}}}\ (\bibinfo
  {publisher} {Springer Netherlands},\ \bibinfo {address} {Dordrecht},\
  \bibinfo {year} {2010})\ pp.\ \bibinfo {pages} {37--68}\BibitemShut {NoStop}%
\bibitem [{\citenamefont {Chevy}(2006)}]{Chevy2006}%
  \BibitemOpen
  \bibfield  {author} {\bibinfo {author} {\bibfnamefont {F.}~\bibnamefont
  {Chevy}},\ }\href {\doibase 10.1103/PhysRevA.74.063628} {\bibfield  {journal}
  {\bibinfo  {journal} {Phys. Rev. A}\ }\textbf {\bibinfo {volume} {74}},\
  \bibinfo {pages} {063628} (\bibinfo {year} {2006})}\BibitemShut {NoStop}%
\bibitem [{\citenamefont {Scazza}\ \emph {et~al.}(2017)\citenamefont {Scazza},
  \citenamefont {Valtolina}, \citenamefont {Massignan}, \citenamefont {Recati},
  \citenamefont {Amico}, \citenamefont {Burchianti}, \citenamefont {Fort},
  \citenamefont {Inguscio}, \citenamefont {Zaccanti},\ and\ \citenamefont
  {Roati}}]{ScazzaValtolina2017}%
  \BibitemOpen
  \bibfield  {author} {\bibinfo {author} {\bibfnamefont {F.}~\bibnamefont
  {Scazza}}, \bibinfo {author} {\bibfnamefont {G.}~\bibnamefont {Valtolina}},
  \bibinfo {author} {\bibfnamefont {P.}~\bibnamefont {Massignan}}, \bibinfo
  {author} {\bibfnamefont {A.}~\bibnamefont {Recati}}, \bibinfo {author}
  {\bibfnamefont {A.}~\bibnamefont {Amico}}, \bibinfo {author} {\bibfnamefont
  {A.}~\bibnamefont {Burchianti}}, \bibinfo {author} {\bibfnamefont
  {C.}~\bibnamefont {Fort}}, \bibinfo {author} {\bibfnamefont {M.}~\bibnamefont
  {Inguscio}}, \bibinfo {author} {\bibfnamefont {M.}~\bibnamefont {Zaccanti}},
  \ and\ \bibinfo {author} {\bibfnamefont {G.}~\bibnamefont {Roati}},\ }\href
  {\doibase 10.1103/PhysRevLett.118.083602} {\bibfield  {journal} {\bibinfo
  {journal} {Phys. Rev. Lett.}\ }\textbf {\bibinfo {volume} {118}},\ \bibinfo
  {pages} {083602} (\bibinfo {year} {2017})}\BibitemShut {NoStop}%
\bibitem [{\citenamefont {Kohstall}\ \emph {et~al.}(2012)\citenamefont
  {Kohstall}, \citenamefont {Zaccanti}, \citenamefont {Jag}, \citenamefont
  {Trenkwalder}, \citenamefont {Massignan}, \citenamefont {Bruun},
  \citenamefont {Schreck},\ and\ \citenamefont {Grimm}}]{KohstallZaccanti2012}%
  \BibitemOpen
  \bibfield  {author} {\bibinfo {author} {\bibfnamefont {C.}~\bibnamefont
  {Kohstall}}, \bibinfo {author} {\bibfnamefont {M.}~\bibnamefont {Zaccanti}},
  \bibinfo {author} {\bibfnamefont {M.}~\bibnamefont {Jag}}, \bibinfo {author}
  {\bibfnamefont {A.}~\bibnamefont {Trenkwalder}}, \bibinfo {author}
  {\bibfnamefont {P.}~\bibnamefont {Massignan}}, \bibinfo {author}
  {\bibfnamefont {G.~M.}\ \bibnamefont {Bruun}}, \bibinfo {author}
  {\bibfnamefont {F.}~\bibnamefont {Schreck}}, \ and\ \bibinfo {author}
  {\bibfnamefont {R.}~\bibnamefont {Grimm}},\ }\href {\doibase
  10.1038/nature11065} {\bibfield  {journal} {\bibinfo  {journal} {Nature}\
  }\textbf {\bibinfo {volume} {485}},\ \bibinfo {pages} {615} (\bibinfo {year}
  {2012})}\BibitemShut {NoStop}%
\bibitem [{\citenamefont {Schirotzek}\ \emph {et~al.}(2009)\citenamefont
  {Schirotzek}, \citenamefont {Wu}, \citenamefont {Sommer},\ and\ \citenamefont
  {Zwierlein}}]{SchirotzekWu2009}%
  \BibitemOpen
  \bibfield  {author} {\bibinfo {author} {\bibfnamefont {A.}~\bibnamefont
  {Schirotzek}}, \bibinfo {author} {\bibfnamefont {C.-H.}\ \bibnamefont {Wu}},
  \bibinfo {author} {\bibfnamefont {A.}~\bibnamefont {Sommer}}, \ and\ \bibinfo
  {author} {\bibfnamefont {M.~W.}\ \bibnamefont {Zwierlein}},\ }\href {\doibase
  10.1103/PhysRevLett.102.230402} {\bibfield  {journal} {\bibinfo  {journal}
  {Phys. Rev. Lett.}\ }\textbf {\bibinfo {volume} {102}},\ \bibinfo {pages}
  {230402} (\bibinfo {year} {2009})}\BibitemShut {NoStop}%
\bibitem [{\citenamefont {Cetina}\ \emph {et~al.}(2016)\citenamefont {Cetina},
  \citenamefont {Jag}, \citenamefont {Lous}, \citenamefont {Fritsche},
  \citenamefont {Walraven}, \citenamefont {Grimm}, \citenamefont {Levinsen},
  \citenamefont {Parish}, \citenamefont {Schmidt}, \citenamefont {Knap},\ and\
  \citenamefont {Demler}}]{CetinaJag2016}%
  \BibitemOpen
  \bibfield  {author} {\bibinfo {author} {\bibfnamefont {M.}~\bibnamefont
  {Cetina}}, \bibinfo {author} {\bibfnamefont {M.}~\bibnamefont {Jag}},
  \bibinfo {author} {\bibfnamefont {R.~S.}\ \bibnamefont {Lous}}, \bibinfo
  {author} {\bibfnamefont {I.}~\bibnamefont {Fritsche}}, \bibinfo {author}
  {\bibfnamefont {J.~T.~M.}\ \bibnamefont {Walraven}}, \bibinfo {author}
  {\bibfnamefont {R.}~\bibnamefont {Grimm}}, \bibinfo {author} {\bibfnamefont
  {J.}~\bibnamefont {Levinsen}}, \bibinfo {author} {\bibfnamefont {M.~M.}\
  \bibnamefont {Parish}}, \bibinfo {author} {\bibfnamefont {R.}~\bibnamefont
  {Schmidt}}, \bibinfo {author} {\bibfnamefont {M.}~\bibnamefont {Knap}}, \
  and\ \bibinfo {author} {\bibfnamefont {E.}~\bibnamefont {Demler}},\ }\href
  {\doibase 10.1126/science.aaf5134} {\bibfield  {journal} {\bibinfo  {journal}
  {Science}\ }\textbf {\bibinfo {volume} {354}},\ \bibinfo {pages} {96}
  (\bibinfo {year} {2016})}\BibitemShut {NoStop}%
\bibitem [{\citenamefont {Dirac}(1930)}]{Dirac1930annihilation}%
  \BibitemOpen
  \bibfield  {author} {\bibinfo {author} {\bibfnamefont {P.~A.~M.}\
  \bibnamefont {Dirac}},\ }\href {\doibase 10.1017/S0305004100016091}
  {\bibfield  {journal} {\bibinfo  {journal} {Proc. Cambridge Phil. Soc.}\
  }\textbf {\bibinfo {volume} {26}},\ \bibinfo {pages} {361–375} (\bibinfo
  {year} {1930})}\BibitemShut {NoStop}%
\bibitem [{\citenamefont {Frenkel}(1934)}]{frenkel1934wave}%
  \BibitemOpen
  \bibfield  {author} {\bibinfo {author} {\bibfnamefont {J.}~\bibnamefont
  {Frenkel}},\ }\href@noop {} {\emph {\bibinfo {title} {Wave mechanics,
  advanced general theory}}},\ Vol.~\bibinfo {volume} {1}\ (\bibinfo
  {publisher} {Oxford University Press},\ \bibinfo {year} {1934})\BibitemShut
  {NoStop}%
\bibitem [{\citenamefont {Beck}\ \emph {et~al.}(2000)\citenamefont {Beck},
  \citenamefont {Jäckle}, \citenamefont {Worth},\ and\ \citenamefont
  {Meyer}}]{BeckJackle2000}%
  \BibitemOpen
  \bibfield  {author} {\bibinfo {author} {\bibfnamefont {M.}~\bibnamefont
  {Beck}}, \bibinfo {author} {\bibfnamefont {A.}~\bibnamefont {Jäckle}},
  \bibinfo {author} {\bibfnamefont {G.}~\bibnamefont {Worth}}, \ and\ \bibinfo
  {author} {\bibfnamefont {H.-D.}\ \bibnamefont {Meyer}},\ }\href {\doibase
  https://doi.org/10.1016/S0370-1573(99)00047-2} {\bibfield  {journal}
  {\bibinfo  {journal} {Phys. Rep.}\ }\textbf {\bibinfo {volume} {324}},\
  \bibinfo {pages} {1} (\bibinfo {year} {2000})}\BibitemShut {NoStop}%
\bibitem [{\citenamefont {Kramer}\ and\ \citenamefont
  {Saraceno}(1981)}]{kramer1981geometry}%
  \BibitemOpen
  \bibfield  {author} {\bibinfo {author} {\bibfnamefont {P.}~\bibnamefont
  {Kramer}}\ and\ \bibinfo {author} {\bibfnamefont {M.}~\bibnamefont
  {Saraceno}},\ }\href {https://books.google.at/books?id=UxYvAAAAIAAJ} {\emph
  {\bibinfo {title} {Geometry of the Time-dependent Variational Principle in
  Quantum Mechanics}}},\ Lecture notes in physics\ (\bibinfo  {publisher}
  {Springer-Verlag},\ \bibinfo {year} {1981})\BibitemShut {NoStop}%
\bibitem [{\citenamefont {Kull}\ and\ \citenamefont
  {Pfirsch}(2000)}]{Kull2000generalized}%
  \BibitemOpen
  \bibfield  {author} {\bibinfo {author} {\bibfnamefont {H.-J.}\ \bibnamefont
  {Kull}}\ and\ \bibinfo {author} {\bibfnamefont {D.}~\bibnamefont {Pfirsch}},\
  }\href {\doibase 10.1103/PhysRevE.61.5940} {\bibfield  {journal} {\bibinfo
  {journal} {Phys. Rev. E}\ }\textbf {\bibinfo {volume} {61}},\ \bibinfo
  {pages} {5940} (\bibinfo {year} {2000})}\BibitemShut {NoStop}%
\bibitem [{\citenamefont {McLachlan}(1964)}]{McLachlan1964variational}%
  \BibitemOpen
  \bibfield  {author} {\bibinfo {author} {\bibfnamefont {A.~D.}\ \bibnamefont
  {McLachlan}},\ }\href {\doibase 10.1080/00268976400100041} {\bibfield
  {journal} {\bibinfo  {journal} {Mol. Phys.}\ }\textbf {\bibinfo {volume}
  {8}},\ \bibinfo {pages} {39} (\bibinfo {year} {1964})}\BibitemShut {NoStop}%
\bibitem [{\citenamefont {Broeckhove}\ \emph {et~al.}(1988)\citenamefont
  {Broeckhove}, \citenamefont {Lathouwers}, \citenamefont {Kesteloot},\ and\
  \citenamefont {{Van Leuven}}}]{Broeckhove1988equivalence}%
  \BibitemOpen
  \bibfield  {author} {\bibinfo {author} {\bibfnamefont {J.}~\bibnamefont
  {Broeckhove}}, \bibinfo {author} {\bibfnamefont {L.}~\bibnamefont
  {Lathouwers}}, \bibinfo {author} {\bibfnamefont {E.}~\bibnamefont
  {Kesteloot}}, \ and\ \bibinfo {author} {\bibfnamefont {P.}~\bibnamefont {{Van
  Leuven}}},\ }\href {\doibase https://doi.org/10.1016/0009-2614(88)80380-4}
  {\bibfield  {journal} {\bibinfo  {journal} {Chem. Phys. Lett.}\ }\textbf
  {\bibinfo {volume} {149}},\ \bibinfo {pages} {547} (\bibinfo {year}
  {1988})}\BibitemShut {NoStop}%
\bibitem [{\citenamefont {Thouless}(1960)}]{Thouless1960stability}%
  \BibitemOpen
  \bibfield  {author} {\bibinfo {author} {\bibfnamefont {D.}~\bibnamefont
  {Thouless}},\ }\href {\doibase https://doi.org/10.1016/0029-5582(60)90048-1}
  {\bibfield  {journal} {\bibinfo  {journal} {Nuc. Phys.}\ }\textbf {\bibinfo
  {volume} {21}},\ \bibinfo {pages} {225} (\bibinfo {year} {1960})}\BibitemShut
  {NoStop}%
\end{thebibliography}%


\begin{thebibliography}{71}%
\makeatletter
\providecommand \@ifxundefined [1]{%
 \@ifx{#1\undefined}
}%
\providecommand \@ifnum [1]{%
 \ifnum #1\expandafter \@firstoftwo
 \else \expandafter \@secondoftwo
 \fi
}%
\providecommand \@ifx [1]{%
 \ifx #1\expandafter \@firstoftwo
 \else \expandafter \@secondoftwo
 \fi
}%
\providecommand \natexlab [1]{#1}%
\providecommand \enquote  [1]{``#1''}%
\providecommand \bibnamefont  [1]{#1}%
\providecommand \bibfnamefont [1]{#1}%
\providecommand \citenamefont [1]{#1}%
\providecommand \href@noop [0]{\@secondoftwo}%
\providecommand \href [0]{\begingroup \@sanitize@url \@href}%
\providecommand \@href[1]{\@@startlink{#1}\@@href}%
\providecommand \@@href[1]{\endgroup#1\@@endlink}%
\providecommand \@sanitize@url [0]{\catcode `\\12\catcode `\$12\catcode
  `\&12\catcode `\#12\catcode `\^12\catcode `\_12\catcode `\%12\relax}%
\providecommand \@@startlink[1]{}%
\providecommand \@@endlink[0]{}%
\providecommand \url  [0]{\begingroup\@sanitize@url \@url }%
\providecommand \@url [1]{\endgroup\@href {#1}{\urlprefix }}%
\providecommand \urlprefix  [0]{URL }%
\providecommand \Eprint [0]{\href }%
\providecommand \doibase [0]{http://dx.doi.org/}%
\providecommand \selectlanguage [0]{\@gobble}%
\providecommand \bibinfo  [0]{\@secondoftwo}%
\providecommand \bibfield  [0]{\@secondoftwo}%
\providecommand \translation [1]{[#1]}%
\providecommand \BibitemOpen [0]{}%
\providecommand \bibitemStop [0]{}%
\providecommand \bibitemNoStop [0]{.\EOS\space}%
\providecommand \EOS [0]{\spacefactor3000\relax}%
\providecommand \BibitemShut  [1]{\csname bibitem#1\endcsname}%
\let\auto@bib@innerbib\@empty
%</preamble>
\bibitem [{\citenamefont {Brenner}\ \emph {et~al.}(2016)\citenamefont
  {Brenner}, \citenamefont {Egger}, \citenamefont {Kronik}, \citenamefont
  {Hodes},\ and\ \citenamefont {Cahen}}]{brenner2016hybrid}%
  \BibitemOpen
  \bibfield  {author} {\bibinfo {author} {\bibfnamefont {T.~M.}\ \bibnamefont
  {Brenner}}, \bibinfo {author} {\bibfnamefont {D.~A.}\ \bibnamefont {Egger}},
  \bibinfo {author} {\bibfnamefont {L.}~\bibnamefont {Kronik}}, \bibinfo
  {author} {\bibfnamefont {G.}~\bibnamefont {Hodes}}, \ and\ \bibinfo {author}
  {\bibfnamefont {D.}~\bibnamefont {Cahen}},\ }\href
  {https://doi.org/10.1038/natrevmats.2015.7} {\bibfield  {journal} {\bibinfo
  {journal} {Nat. Rev. Mater.}\ }\textbf {\bibinfo {volume} {1}},\ \bibinfo
  {pages} {1} (\bibinfo {year} {2016})}\BibitemShut {NoStop}%
\bibitem [{\citenamefont {Johnston}\ and\ \citenamefont
  {Herz}(2016)}]{johnston2016hybrid}%
  \BibitemOpen
  \bibfield  {author} {\bibinfo {author} {\bibfnamefont {M.~B.}\ \bibnamefont
  {Johnston}}\ and\ \bibinfo {author} {\bibfnamefont {L.~M.}\ \bibnamefont
  {Herz}},\ }\href {https://doi.org/10.1021/acs.accounts.5b00411} {\bibfield
  {journal} {\bibinfo  {journal} {Acc. Chem. Res.}\ }\textbf {\bibinfo {volume}
  {49}},\ \bibinfo {pages} {146} (\bibinfo {year} {2016})}\BibitemShut
  {NoStop}%
\bibitem [{\citenamefont {Jena}\ \emph {et~al.}(2019)\citenamefont {Jena},
  \citenamefont {Kulkarni},\ and\ \citenamefont {Miyasaka}}]{jena2019halide}%
  \BibitemOpen
  \bibfield  {author} {\bibinfo {author} {\bibfnamefont {A.~K.}\ \bibnamefont
  {Jena}}, \bibinfo {author} {\bibfnamefont {A.}~\bibnamefont {Kulkarni}}, \
  and\ \bibinfo {author} {\bibfnamefont {T.}~\bibnamefont {Miyasaka}},\ }\href
  {https://doi.org/10.1021/acs.chemrev.8b00539} {\bibfield  {journal} {\bibinfo
   {journal} {Chem. Rev.}\ }\textbf {\bibinfo {volume} {119}},\ \bibinfo
  {pages} {3036} (\bibinfo {year} {2019})}\BibitemShut {NoStop}%
\bibitem [{\citenamefont {Kojima}\ \emph {et~al.}(2009)\citenamefont {Kojima},
  \citenamefont {Teshima}, \citenamefont {Shirai},\ and\ \citenamefont
  {Miyasaka}}]{kojima2009organometal}%
  \BibitemOpen
  \bibfield  {author} {\bibinfo {author} {\bibfnamefont {A.}~\bibnamefont
  {Kojima}}, \bibinfo {author} {\bibfnamefont {K.}~\bibnamefont {Teshima}},
  \bibinfo {author} {\bibfnamefont {Y.}~\bibnamefont {Shirai}}, \ and\ \bibinfo
  {author} {\bibfnamefont {T.}~\bibnamefont {Miyasaka}},\ }\href
  {https://doi.org/10.1021/ja809598r} {\bibfield  {journal} {\bibinfo
  {journal} {J. Am. Chem. Soc.}\ }\textbf {\bibinfo {volume} {131}},\ \bibinfo
  {pages} {6050} (\bibinfo {year} {2009})}\BibitemShut {NoStop}%
\bibitem [{\citenamefont {Chung}\ \emph {et~al.}(2012)\citenamefont {Chung},
  \citenamefont {Lee}, \citenamefont {He}, \citenamefont {Chang},\ and\
  \citenamefont {Kanatzidis}}]{chung2012all}%
  \BibitemOpen
  \bibfield  {author} {\bibinfo {author} {\bibfnamefont {I.}~\bibnamefont
  {Chung}}, \bibinfo {author} {\bibfnamefont {B.}~\bibnamefont {Lee}}, \bibinfo
  {author} {\bibfnamefont {J.}~\bibnamefont {He}}, \bibinfo {author}
  {\bibfnamefont {R.~P.}\ \bibnamefont {Chang}}, \ and\ \bibinfo {author}
  {\bibfnamefont {M.~G.}\ \bibnamefont {Kanatzidis}},\ }\href
  {https://doi.org/10.1038/nature11067} {\bibfield  {journal} {\bibinfo
  {journal} {Nature}\ }\textbf {\bibinfo {volume} {485}},\ \bibinfo {pages}
  {486} (\bibinfo {year} {2012})}\BibitemShut {NoStop}%
\bibitem [{\citenamefont {Lee}\ \emph {et~al.}(2012)\citenamefont {Lee},
  \citenamefont {Teuscher}, \citenamefont {Miyasaka}, \citenamefont
  {Murakami},\ and\ \citenamefont {Snaith}}]{lee2012efficient}%
  \BibitemOpen
  \bibfield  {author} {\bibinfo {author} {\bibfnamefont {M.~M.}\ \bibnamefont
  {Lee}}, \bibinfo {author} {\bibfnamefont {J.}~\bibnamefont {Teuscher}},
  \bibinfo {author} {\bibfnamefont {T.}~\bibnamefont {Miyasaka}}, \bibinfo
  {author} {\bibfnamefont {T.~N.}\ \bibnamefont {Murakami}}, \ and\ \bibinfo
  {author} {\bibfnamefont {H.~J.}\ \bibnamefont {Snaith}},\ }\href
  {https://doi.org/10.1126/science.1228604} {\bibfield  {journal} {\bibinfo
  {journal} {Science}\ }\textbf {\bibinfo {volume} {338}},\ \bibinfo {pages}
  {643} (\bibinfo {year} {2012})}\BibitemShut {NoStop}%
\bibitem [{\citenamefont {Kim}\ \emph {et~al.}(2012)\citenamefont {Kim},
  \citenamefont {Lee}, \citenamefont {Im}, \citenamefont {Lee}, \citenamefont
  {Moehl}, \citenamefont {Marchioro}, \citenamefont {Moon}, \citenamefont
  {Humphry-Baker}, \citenamefont {Yum}, \citenamefont {Moser}, \citenamefont
  {Grätzel},\ and\ \citenamefont {Park}}]{kim2012lead}%
  \BibitemOpen
  \bibfield  {author} {\bibinfo {author} {\bibfnamefont {H.-S.}\ \bibnamefont
  {Kim}}, \bibinfo {author} {\bibfnamefont {C.-R.}\ \bibnamefont {Lee}},
  \bibinfo {author} {\bibfnamefont {J.-H.}\ \bibnamefont {Im}}, \bibinfo
  {author} {\bibfnamefont {K.-B.}\ \bibnamefont {Lee}}, \bibinfo {author}
  {\bibfnamefont {T.}~\bibnamefont {Moehl}}, \bibinfo {author} {\bibfnamefont
  {A.}~\bibnamefont {Marchioro}}, \bibinfo {author} {\bibfnamefont {S.-J.}\
  \bibnamefont {Moon}}, \bibinfo {author} {\bibfnamefont {R.}~\bibnamefont
  {Humphry-Baker}}, \bibinfo {author} {\bibfnamefont {J.-H.}\ \bibnamefont
  {Yum}}, \bibinfo {author} {\bibfnamefont {J.~E.}\ \bibnamefont {Moser}},
  \bibinfo {author} {\bibfnamefont {M.}~\bibnamefont {Grätzel}}, \ and\
  \bibinfo {author} {\bibfnamefont {N.-G.}\ \bibnamefont {Park}},\ }\href
  {https://doi.org/10.1038/srep00591} {\bibfield  {journal} {\bibinfo
  {journal} {Sci. Rep.}\ }\textbf {\bibinfo {volume} {2}},\ \bibinfo {pages}
  {1} (\bibinfo {year} {2012})}\BibitemShut {NoStop}%
\bibitem [{\citenamefont {Miyasaka}(2021)}]{miyasaka2021perovskite}%
  \BibitemOpen
  \bibfield  {author} {\bibinfo {author} {\bibfnamefont {T.}~\bibnamefont
  {Miyasaka}},\ }\href@noop {} {\emph {\bibinfo {title} {Perovskite
  Photovoltaics and Optoelectronics: From Fundamentals to Advanced
  Applications}}}\ (\bibinfo  {publisher} {John Wiley \& Sons},\ \bibinfo
  {year} {2021})\BibitemShut {NoStop}%
\bibitem [{\citenamefont {Fujiwara}(2022)}]{fujiwara2022hybrid}%
  \BibitemOpen
  \bibfield  {author} {\bibinfo {author} {\bibfnamefont {H.}~\bibnamefont
  {Fujiwara}},\ }\href@noop {} {\emph {\bibinfo {title} {Hybrid Perovskite
  Solar Cells: Characteristics and Operation}}}\ (\bibinfo  {publisher} {John
  Wiley \& Sons},\ \bibinfo {year} {2022})\BibitemShut {NoStop}%
\bibitem [{\citenamefont {Miyata}\ \emph
  {et~al.}(2017{\natexlab{a}})\citenamefont {Miyata}, \citenamefont {Atallah},\
  and\ \citenamefont {Zhu}}]{miyata2017lead}%
  \BibitemOpen
  \bibfield  {author} {\bibinfo {author} {\bibfnamefont {K.}~\bibnamefont
  {Miyata}}, \bibinfo {author} {\bibfnamefont {T.~L.}\ \bibnamefont {Atallah}},
  \ and\ \bibinfo {author} {\bibfnamefont {X.-Y.}\ \bibnamefont {Zhu}},\ }\href
  {https://doi.org/10.1126/sciadv.1701469} {\bibfield  {journal} {\bibinfo
  {journal} {Sci. Adv.}\ }\textbf {\bibinfo {volume} {3}},\ \bibinfo {pages}
  {e1701469} (\bibinfo {year} {2017}{\natexlab{a}})}\BibitemShut {NoStop}%
\bibitem [{\citenamefont {Eames}\ \emph {et~al.}(2015)\citenamefont {Eames},
  \citenamefont {Frost}, \citenamefont {Barnes}, \citenamefont {O’regan},
  \citenamefont {Walsh},\ and\ \citenamefont {Islam}}]{eames2015ionic}%
  \BibitemOpen
  \bibfield  {author} {\bibinfo {author} {\bibfnamefont {C.}~\bibnamefont
  {Eames}}, \bibinfo {author} {\bibfnamefont {J.~M.}\ \bibnamefont {Frost}},
  \bibinfo {author} {\bibfnamefont {P.~R.}\ \bibnamefont {Barnes}}, \bibinfo
  {author} {\bibfnamefont {B.~C.}\ \bibnamefont {O’regan}}, \bibinfo {author}
  {\bibfnamefont {A.}~\bibnamefont {Walsh}}, \ and\ \bibinfo {author}
  {\bibfnamefont {M.~S.}\ \bibnamefont {Islam}},\ }\href
  {https://doi.org/10.1038/ncomms8497} {\bibfield  {journal} {\bibinfo
  {journal} {Nat. Commun.}\ }\textbf {\bibinfo {volume} {6}},\ \bibinfo {pages}
  {1} (\bibinfo {year} {2015})}\BibitemShut {NoStop}%
\bibitem [{\citenamefont {Yuan}\ and\ \citenamefont
  {Huang}(2016)}]{yuan2016ion}%
  \BibitemOpen
  \bibfield  {author} {\bibinfo {author} {\bibfnamefont {Y.}~\bibnamefont
  {Yuan}}\ and\ \bibinfo {author} {\bibfnamefont {J.}~\bibnamefont {Huang}},\
  }\href {https://doi.org/10.1021/acs.accounts.5b00420} {\bibfield  {journal}
  {\bibinfo  {journal} {Acc. Chem. Res.}\ }\textbf {\bibinfo {volume} {49}},\
  \bibinfo {pages} {286} (\bibinfo {year} {2016})}\BibitemShut {NoStop}%
\bibitem [{\citenamefont {Chen}\ \emph {et~al.}(2017)\citenamefont {Chen},
  \citenamefont {Chen}, \citenamefont {Foley}, \citenamefont {Lee},
  \citenamefont {Ruff}, \citenamefont {Ko}, \citenamefont {Brown},
  \citenamefont {Harriger}, \citenamefont {Zhang}, \citenamefont {Park},
  \citenamefont {Yoon}, \citenamefont {Chang}, \citenamefont {Choi},\ and\
  \citenamefont {Lee}}]{chen2017origin}%
  \BibitemOpen
  \bibfield  {author} {\bibinfo {author} {\bibfnamefont {T.}~\bibnamefont
  {Chen}}, \bibinfo {author} {\bibfnamefont {W.-L.}\ \bibnamefont {Chen}},
  \bibinfo {author} {\bibfnamefont {B.~J.}\ \bibnamefont {Foley}}, \bibinfo
  {author} {\bibfnamefont {J.}~\bibnamefont {Lee}}, \bibinfo {author}
  {\bibfnamefont {J.~P.}\ \bibnamefont {Ruff}}, \bibinfo {author}
  {\bibfnamefont {J.~P.}\ \bibnamefont {Ko}}, \bibinfo {author} {\bibfnamefont
  {C.~M.}\ \bibnamefont {Brown}}, \bibinfo {author} {\bibfnamefont {L.~W.}\
  \bibnamefont {Harriger}}, \bibinfo {author} {\bibfnamefont {D.}~\bibnamefont
  {Zhang}}, \bibinfo {author} {\bibfnamefont {C.}~\bibnamefont {Park}},
  \bibinfo {author} {\bibfnamefont {M.}~\bibnamefont {Yoon}}, \bibinfo {author}
  {\bibfnamefont {Y.-M.}\ \bibnamefont {Chang}}, \bibinfo {author}
  {\bibfnamefont {J.~J.}\ \bibnamefont {Choi}}, \ and\ \bibinfo {author}
  {\bibfnamefont {S.-H.}\ \bibnamefont {Lee}},\ }\href
  {https://doi.org/10.1073/pnas.1704421114} {\bibfield  {journal} {\bibinfo
  {journal} {Proc. Natl. Acad. Sci. U.S.A.}\ }\textbf {\bibinfo {volume}
  {114}},\ \bibinfo {pages} {7519} (\bibinfo {year} {2017})}\BibitemShut
  {NoStop}%
\bibitem [{\citenamefont {Selig}\ \emph {et~al.}(2017)\citenamefont {Selig},
  \citenamefont {Sadhanala}, \citenamefont {M{\" u}ller}, \citenamefont
  {Lovrincic}, \citenamefont {Chen}, \citenamefont {Rezus}, \citenamefont
  {Frost}, \citenamefont {Jansen},\ and\ \citenamefont
  {Bakulin}}]{selig2017organic}%
  \BibitemOpen
  \bibfield  {author} {\bibinfo {author} {\bibfnamefont {O.}~\bibnamefont
  {Selig}}, \bibinfo {author} {\bibfnamefont {A.}~\bibnamefont {Sadhanala}},
  \bibinfo {author} {\bibfnamefont {C.}~\bibnamefont {M{\" u}ller}}, \bibinfo
  {author} {\bibfnamefont {R.}~\bibnamefont {Lovrincic}}, \bibinfo {author}
  {\bibfnamefont {Z.}~\bibnamefont {Chen}}, \bibinfo {author} {\bibfnamefont
  {Y.~L.}\ \bibnamefont {Rezus}}, \bibinfo {author} {\bibfnamefont {J.~M.}\
  \bibnamefont {Frost}}, \bibinfo {author} {\bibfnamefont {T.~L.}\ \bibnamefont
  {Jansen}}, \ and\ \bibinfo {author} {\bibfnamefont {A.~A.}\ \bibnamefont
  {Bakulin}},\ }\href {https://doi.org/10.1021/jacs.6b12239} {\bibfield
  {journal} {\bibinfo  {journal} {J. Am. Chem. Soc.}\ }\textbf {\bibinfo
  {volume} {139}},\ \bibinfo {pages} {4068} (\bibinfo {year}
  {2017})}\BibitemShut {NoStop}%
\bibitem [{\citenamefont {Liu}\ \emph {et~al.}(2022)\citenamefont {Liu},
  \citenamefont {Guo},\ and\ \citenamefont {Xie}}]{liu2022effects}%
  \BibitemOpen
  \bibfield  {author} {\bibinfo {author} {\bibfnamefont {S.}~\bibnamefont
  {Liu}}, \bibinfo {author} {\bibfnamefont {R.}~\bibnamefont {Guo}}, \ and\
  \bibinfo {author} {\bibfnamefont {F.}~\bibnamefont {Xie}},\ }\href
  {https://doi.org/10.1016/j.matdes.2022.110951} {\bibfield  {journal}
  {\bibinfo  {journal} {Mater. Des.}\ }\textbf {\bibinfo {volume} {221}},\
  \bibinfo {pages} {110951} (\bibinfo {year} {2022})}\BibitemShut {NoStop}%
\bibitem [{\citenamefont {Mozur}\ and\ \citenamefont
  {Neilson}(2021)}]{mozur2021cation}%
  \BibitemOpen
  \bibfield  {author} {\bibinfo {author} {\bibfnamefont {E.~M.}\ \bibnamefont
  {Mozur}}\ and\ \bibinfo {author} {\bibfnamefont {J.~R.}\ \bibnamefont
  {Neilson}},\ }\href {https://doi.org/10.1146/annurev-matsci-080819-012808}
  {\bibfield  {journal} {\bibinfo  {journal} {Annu. Rev. Mater. Res.}\ }\textbf
  {\bibinfo {volume} {51}},\ \bibinfo {pages} {269} (\bibinfo {year}
  {2021})}\BibitemShut {NoStop}%
\bibitem [{\citenamefont {Even}\ \emph
  {et~al.}(2014{\natexlab{a}})\citenamefont {Even}, \citenamefont {Pedesseau},
  \citenamefont {Jancu},\ and\ \citenamefont {Katan}}]{even2014dft}%
  \BibitemOpen
  \bibfield  {author} {\bibinfo {author} {\bibfnamefont {J.}~\bibnamefont
  {Even}}, \bibinfo {author} {\bibfnamefont {L.}~\bibnamefont {Pedesseau}},
  \bibinfo {author} {\bibfnamefont {J.-M.}\ \bibnamefont {Jancu}}, \ and\
  \bibinfo {author} {\bibfnamefont {C.}~\bibnamefont {Katan}},\ }\href
  {https://doi.org/10.1002/pssr.201308183} {\bibfield  {journal} {\bibinfo
  {journal} {Phys. Status Solidi RRL}\ }\textbf {\bibinfo {volume} {8}},\
  \bibinfo {pages} {31} (\bibinfo {year} {2014}{\natexlab{a}})}\BibitemShut
  {NoStop}%
\bibitem [{\citenamefont {Even}\ \emph {et~al.}(2015)\citenamefont {Even},
  \citenamefont {Pedesseau}, \citenamefont {Katan}, \citenamefont {Kepenekian},
  \citenamefont {Lauret}, \citenamefont {Sapori},\ and\ \citenamefont
  {Deleporte}}]{even2015solid}%
  \BibitemOpen
  \bibfield  {author} {\bibinfo {author} {\bibfnamefont {J.}~\bibnamefont
  {Even}}, \bibinfo {author} {\bibfnamefont {L.}~\bibnamefont {Pedesseau}},
  \bibinfo {author} {\bibfnamefont {C.}~\bibnamefont {Katan}}, \bibinfo
  {author} {\bibfnamefont {M.}~\bibnamefont {Kepenekian}}, \bibinfo {author}
  {\bibfnamefont {J.-S.}\ \bibnamefont {Lauret}}, \bibinfo {author}
  {\bibfnamefont {D.}~\bibnamefont {Sapori}}, \ and\ \bibinfo {author}
  {\bibfnamefont {E.}~\bibnamefont {Deleporte}},\ }\href
  {https://doi.org/10.1021/acs.jpcc.5b00695} {\bibfield  {journal} {\bibinfo
  {journal} {J. Phys. Chem. C}\ }\textbf {\bibinfo {volume} {119}},\ \bibinfo
  {pages} {10161} (\bibinfo {year} {2015})}\BibitemShut {NoStop}%
\bibitem [{\citenamefont {Yun}\ \emph {et~al.}(2017)\citenamefont {Yun},
  \citenamefont {Zhou}, \citenamefont {Even},\ and\ \citenamefont
  {Hagfeldt}}]{yun2017theoretical}%
  \BibitemOpen
  \bibfield  {author} {\bibinfo {author} {\bibfnamefont {S.}~\bibnamefont
  {Yun}}, \bibinfo {author} {\bibfnamefont {X.}~\bibnamefont {Zhou}}, \bibinfo
  {author} {\bibfnamefont {J.}~\bibnamefont {Even}}, \ and\ \bibinfo {author}
  {\bibfnamefont {A.}~\bibnamefont {Hagfeldt}},\ }\href
  {https://doi.org/10.1002/anie.201702660} {\bibfield  {journal} {\bibinfo
  {journal} {Angew. Chem. Int. Ed. Engl.}\ }\textbf {\bibinfo {volume} {56}},\
  \bibinfo {pages} {15806} (\bibinfo {year} {2017})}\BibitemShut {NoStop}%
\bibitem [{\citenamefont {Traor{\'e}}\ \emph {et~al.}(2022)\citenamefont
  {Traor{\'e}}, \citenamefont {Even}, \citenamefont {Pedesseau}, \citenamefont
  {Kepenekian},\ and\ \citenamefont {Katan}}]{traore2022band}%
  \BibitemOpen
  \bibfield  {author} {\bibinfo {author} {\bibfnamefont {B.}~\bibnamefont
  {Traor{\'e}}}, \bibinfo {author} {\bibfnamefont {J.}~\bibnamefont {Even}},
  \bibinfo {author} {\bibfnamefont {L.}~\bibnamefont {Pedesseau}}, \bibinfo
  {author} {\bibfnamefont {M.}~\bibnamefont {Kepenekian}}, \ and\ \bibinfo
  {author} {\bibfnamefont {C.}~\bibnamefont {Katan}},\ }\href
  {https://doi.org/10.1103/PhysRevMaterials.6.014604} {\bibfield  {journal}
  {\bibinfo  {journal} {Phys. Rev. Mater.}\ }\textbf {\bibinfo {volume} {6}},\
  \bibinfo {pages} {014604} (\bibinfo {year} {2022})}\BibitemShut {NoStop}%
\bibitem [{\citenamefont {Meggiolaro}\ \emph {et~al.}(2020)\citenamefont
  {Meggiolaro}, \citenamefont {Ambrosio}, \citenamefont {Mosconi},
  \citenamefont {Mahata},\ and\ \citenamefont
  {De~Angelis}}]{meggiolaro2020polarons}%
  \BibitemOpen
  \bibfield  {author} {\bibinfo {author} {\bibfnamefont {D.}~\bibnamefont
  {Meggiolaro}}, \bibinfo {author} {\bibfnamefont {F.}~\bibnamefont
  {Ambrosio}}, \bibinfo {author} {\bibfnamefont {E.}~\bibnamefont {Mosconi}},
  \bibinfo {author} {\bibfnamefont {A.}~\bibnamefont {Mahata}}, \ and\ \bibinfo
  {author} {\bibfnamefont {F.}~\bibnamefont {De~Angelis}},\ }\href
  {https://doi.org/10.1002/aenm.201902748} {\bibfield  {journal} {\bibinfo
  {journal} {Adv. Energy Mater.}\ }\textbf {\bibinfo {volume} {10}},\ \bibinfo
  {pages} {1902748} (\bibinfo {year} {2020})}\BibitemShut {NoStop}%
\bibitem [{\citenamefont {Jinnouchi}\ \emph {et~al.}(2019)\citenamefont
  {Jinnouchi}, \citenamefont {Lahnsteiner}, \citenamefont {Karsai},
  \citenamefont {Kresse},\ and\ \citenamefont {Bokdam}}]{jinnouchi2019phase}%
  \BibitemOpen
  \bibfield  {author} {\bibinfo {author} {\bibfnamefont {R.}~\bibnamefont
  {Jinnouchi}}, \bibinfo {author} {\bibfnamefont {J.}~\bibnamefont
  {Lahnsteiner}}, \bibinfo {author} {\bibfnamefont {F.}~\bibnamefont {Karsai}},
  \bibinfo {author} {\bibfnamefont {G.}~\bibnamefont {Kresse}}, \ and\ \bibinfo
  {author} {\bibfnamefont {M.}~\bibnamefont {Bokdam}},\ }\href
  {https://doi.org/10.1103/PhysRevLett.122.225701} {\bibfield  {journal}
  {\bibinfo  {journal} {Phys. Rev. Lett.}\ }\textbf {\bibinfo {volume} {122}},\
  \bibinfo {pages} {225701} (\bibinfo {year} {2019})}\BibitemShut {NoStop}%
\bibitem [{\citenamefont {Zhang}\ \emph {et~al.}(2020)\citenamefont {Zhang},
  \citenamefont {He},\ and\ \citenamefont {Shao}}]{zhang2020machine}%
  \BibitemOpen
  \bibfield  {author} {\bibinfo {author} {\bibfnamefont {L.}~\bibnamefont
  {Zhang}}, \bibinfo {author} {\bibfnamefont {M.}~\bibnamefont {He}}, \ and\
  \bibinfo {author} {\bibfnamefont {S.}~\bibnamefont {Shao}},\ }\href
  {https://doi.org/10.1016/j.nanoen.2020.105380} {\bibfield  {journal}
  {\bibinfo  {journal} {Nano Energy}\ }\textbf {\bibinfo {volume} {78}},\
  \bibinfo {pages} {105380} (\bibinfo {year} {2020})}\BibitemShut {NoStop}%
\bibitem [{\citenamefont {Myung}\ \emph {et~al.}(2022)\citenamefont {Myung},
  \citenamefont {Hajibabaei}, \citenamefont {Cha}, \citenamefont {Ha},
  \citenamefont {Kim},\ and\ \citenamefont {Kim}}]{myung2022challenges}%
  \BibitemOpen
  \bibfield  {author} {\bibinfo {author} {\bibfnamefont {C.~W.}\ \bibnamefont
  {Myung}}, \bibinfo {author} {\bibfnamefont {A.}~\bibnamefont {Hajibabaei}},
  \bibinfo {author} {\bibfnamefont {J.-H.}\ \bibnamefont {Cha}}, \bibinfo
  {author} {\bibfnamefont {M.}~\bibnamefont {Ha}}, \bibinfo {author}
  {\bibfnamefont {J.}~\bibnamefont {Kim}}, \ and\ \bibinfo {author}
  {\bibfnamefont {K.~S.}\ \bibnamefont {Kim}},\ }\href
  {https://doi.org/10.1002/aenm.202202279} {\bibfield  {journal} {\bibinfo
  {journal} {Adv. Energy Mater.}\ }\textbf {\bibinfo {volume} {12}},\ \bibinfo
  {pages} {2202279} (\bibinfo {year} {2022})}\BibitemShut {NoStop}%
\bibitem [{\citenamefont {Brenner}\ \emph {et~al.}(2015)\citenamefont
  {Brenner}, \citenamefont {Egger}, \citenamefont {Rappe}, \citenamefont
  {Kronik}, \citenamefont {Hodes},\ and\ \citenamefont
  {Cahen}}]{brenner2015mobilities}%
  \BibitemOpen
  \bibfield  {author} {\bibinfo {author} {\bibfnamefont {T.~M.}\ \bibnamefont
  {Brenner}}, \bibinfo {author} {\bibfnamefont {D.~A.}\ \bibnamefont {Egger}},
  \bibinfo {author} {\bibfnamefont {A.~M.}\ \bibnamefont {Rappe}}, \bibinfo
  {author} {\bibfnamefont {L.}~\bibnamefont {Kronik}}, \bibinfo {author}
  {\bibfnamefont {G.}~\bibnamefont {Hodes}}, \ and\ \bibinfo {author}
  {\bibfnamefont {D.}~\bibnamefont {Cahen}},\ }\href
  {https://doi.org/10.1021/acs.jpclett.5b02390} {\bibfield  {journal} {\bibinfo
   {journal} {J. Phys. Chem. Lett.}\ }\textbf {\bibinfo {volume} {6}},\
  \bibinfo {pages} {4754} (\bibinfo {year} {2015})}\BibitemShut {NoStop}%
\bibitem [{\citenamefont {Even}\ \emph
  {et~al.}(2014{\natexlab{b}})\citenamefont {Even}, \citenamefont {Pedesseau},\
  and\ \citenamefont {Katan}}]{even2014analysis}%
  \BibitemOpen
  \bibfield  {author} {\bibinfo {author} {\bibfnamefont {J.}~\bibnamefont
  {Even}}, \bibinfo {author} {\bibfnamefont {L.}~\bibnamefont {Pedesseau}}, \
  and\ \bibinfo {author} {\bibfnamefont {C.}~\bibnamefont {Katan}},\ }\href
  {https://doi.org/10.1021/jp503337a} {\bibfield  {journal} {\bibinfo
  {journal} {J. Phys. Chem. C}\ }\textbf {\bibinfo {volume} {118}},\ \bibinfo
  {pages} {11566} (\bibinfo {year} {2014}{\natexlab{b}})}\BibitemShut {NoStop}%
\bibitem [{\citenamefont {Amat}\ \emph {et~al.}(2014)\citenamefont {Amat},
  \citenamefont {Mosconi}, \citenamefont {Ronca}, \citenamefont {Quarti},
  \citenamefont {Umari}, \citenamefont {Nazeeruddin}, \citenamefont {Gratzel},\
  and\ \citenamefont {De~Angelis}}]{amat2014cation}%
  \BibitemOpen
  \bibfield  {author} {\bibinfo {author} {\bibfnamefont {A.}~\bibnamefont
  {Amat}}, \bibinfo {author} {\bibfnamefont {E.}~\bibnamefont {Mosconi}},
  \bibinfo {author} {\bibfnamefont {E.}~\bibnamefont {Ronca}}, \bibinfo
  {author} {\bibfnamefont {C.}~\bibnamefont {Quarti}}, \bibinfo {author}
  {\bibfnamefont {P.}~\bibnamefont {Umari}}, \bibinfo {author} {\bibfnamefont
  {M.~K.}\ \bibnamefont {Nazeeruddin}}, \bibinfo {author} {\bibfnamefont
  {M.}~\bibnamefont {Gratzel}}, \ and\ \bibinfo {author} {\bibfnamefont
  {F.}~\bibnamefont {De~Angelis}},\ }\href {https://doi.org/10.1021/nl5012992}
  {\bibfield  {journal} {\bibinfo  {journal} {Nano Lett.}\ }\textbf {\bibinfo
  {volume} {14}},\ \bibinfo {pages} {3608} (\bibinfo {year}
  {2014})}\BibitemShut {NoStop}%
\bibitem [{\citenamefont {Zheng}\ \emph {et~al.}(2015)\citenamefont {Zheng},
  \citenamefont {Tan}, \citenamefont {Liu},\ and\ \citenamefont
  {Rappe}}]{zheng2015rashba}%
  \BibitemOpen
  \bibfield  {author} {\bibinfo {author} {\bibfnamefont {F.}~\bibnamefont
  {Zheng}}, \bibinfo {author} {\bibfnamefont {L.~Z.}\ \bibnamefont {Tan}},
  \bibinfo {author} {\bibfnamefont {S.}~\bibnamefont {Liu}}, \ and\ \bibinfo
  {author} {\bibfnamefont {A.~M.}\ \bibnamefont {Rappe}},\ }\href
  {https://doi.org/10.1021/acs.nanolett.5b01854} {\bibfield  {journal}
  {\bibinfo  {journal} {Nano Lett.}\ }\textbf {\bibinfo {volume} {15}},\
  \bibinfo {pages} {7794} (\bibinfo {year} {2015})}\BibitemShut {NoStop}%
\bibitem [{\citenamefont {Zhu}\ and\ \citenamefont
  {Podzorov}(2015)}]{zhu2015charge}%
  \BibitemOpen
  \bibfield  {author} {\bibinfo {author} {\bibfnamefont {X.-Y.}\ \bibnamefont
  {Zhu}}\ and\ \bibinfo {author} {\bibfnamefont {V.}~\bibnamefont {Podzorov}},\
  }\href {https://doi.org/10.1021/acs.jpclett.5b02462} {\bibfield  {journal}
  {\bibinfo  {journal} {J. Phys. Chem. Lett.}\ }\textbf {\bibinfo {volume}
  {6}},\ \bibinfo {pages} {4758} (\bibinfo {year} {2015})}\BibitemShut
  {NoStop}%
\bibitem [{\citenamefont {Welch}\ \emph {et~al.}(2016)\citenamefont {Welch},
  \citenamefont {Scolfaro},\ and\ \citenamefont {Zakhidov}}]{welch2016density}%
  \BibitemOpen
  \bibfield  {author} {\bibinfo {author} {\bibfnamefont {E.}~\bibnamefont
  {Welch}}, \bibinfo {author} {\bibfnamefont {L.}~\bibnamefont {Scolfaro}}, \
  and\ \bibinfo {author} {\bibfnamefont {A.}~\bibnamefont {Zakhidov}},\ }\href
  {https://doi.org/10.1063/1.4972341} {\bibfield  {journal} {\bibinfo
  {journal} {AIP Adv.}\ }\textbf {\bibinfo {volume} {6}},\ \bibinfo {pages}
  {125037} (\bibinfo {year} {2016})}\BibitemShut {NoStop}%
\bibitem [{\citenamefont {Neukirch}\ \emph {et~al.}(2016)\citenamefont
  {Neukirch}, \citenamefont {Nie}, \citenamefont {Blancon}, \citenamefont
  {Appavoo}, \citenamefont {Tsai}, \citenamefont {Sfeir}, \citenamefont
  {Katan}, \citenamefont {Pedesseau}, \citenamefont {Even}, \citenamefont
  {Crochet}, \citenamefont {Gupta}, \citenamefont {Mohite},\ and\ \citenamefont
  {Tretiak}}]{neukirch2016polaron}%
  \BibitemOpen
  \bibfield  {author} {\bibinfo {author} {\bibfnamefont {A.~J.}\ \bibnamefont
  {Neukirch}}, \bibinfo {author} {\bibfnamefont {W.}~\bibnamefont {Nie}},
  \bibinfo {author} {\bibfnamefont {J.-C.}\ \bibnamefont {Blancon}}, \bibinfo
  {author} {\bibfnamefont {K.}~\bibnamefont {Appavoo}}, \bibinfo {author}
  {\bibfnamefont {H.}~\bibnamefont {Tsai}}, \bibinfo {author} {\bibfnamefont
  {M.~Y.}\ \bibnamefont {Sfeir}}, \bibinfo {author} {\bibfnamefont
  {C.}~\bibnamefont {Katan}}, \bibinfo {author} {\bibfnamefont
  {L.}~\bibnamefont {Pedesseau}}, \bibinfo {author} {\bibfnamefont
  {J.}~\bibnamefont {Even}}, \bibinfo {author} {\bibfnamefont {J.~J.}\
  \bibnamefont {Crochet}}, \bibinfo {author} {\bibfnamefont {G.}~\bibnamefont
  {Gupta}}, \bibinfo {author} {\bibfnamefont {A.~D.}\ \bibnamefont {Mohite}}, \
  and\ \bibinfo {author} {\bibfnamefont {S.}~\bibnamefont {Tretiak}},\ }\href
  {https://doi.org/10.1021/acs.nanolett.6b01218} {\bibfield  {journal}
  {\bibinfo  {journal} {Nano Lett.}\ }\textbf {\bibinfo {volume} {16}},\
  \bibinfo {pages} {3809} (\bibinfo {year} {2016})}\BibitemShut {NoStop}%
\bibitem [{\citenamefont {Ivanovska}\ \emph {et~al.}(2017)\citenamefont
  {Ivanovska}, \citenamefont {Dionigi}, \citenamefont {Mosconi}, \citenamefont
  {De~Angelis}, \citenamefont {Liscio}, \citenamefont {Morandi},\ and\
  \citenamefont {Ruani}}]{ivanovska2017long}%
  \BibitemOpen
  \bibfield  {author} {\bibinfo {author} {\bibfnamefont {T.}~\bibnamefont
  {Ivanovska}}, \bibinfo {author} {\bibfnamefont {C.}~\bibnamefont {Dionigi}},
  \bibinfo {author} {\bibfnamefont {E.}~\bibnamefont {Mosconi}}, \bibinfo
  {author} {\bibfnamefont {F.}~\bibnamefont {De~Angelis}}, \bibinfo {author}
  {\bibfnamefont {F.}~\bibnamefont {Liscio}}, \bibinfo {author} {\bibfnamefont
  {V.}~\bibnamefont {Morandi}}, \ and\ \bibinfo {author} {\bibfnamefont
  {G.}~\bibnamefont {Ruani}},\ }\href
  {https://doi.org/10.1021/acs.jpclett.7b01156} {\bibfield  {journal} {\bibinfo
   {journal} {J. Phys. Chem. Lett.}\ }\textbf {\bibinfo {volume} {8}},\
  \bibinfo {pages} {3081} (\bibinfo {year} {2017})}\BibitemShut {NoStop}%
\bibitem [{\citenamefont {Zheng}\ and\ \citenamefont
  {Wang}(2019)}]{zheng2019large}%
  \BibitemOpen
  \bibfield  {author} {\bibinfo {author} {\bibfnamefont {F.}~\bibnamefont
  {Zheng}}\ and\ \bibinfo {author} {\bibfnamefont {L.-w.}\ \bibnamefont
  {Wang}},\ }\href {https://doi.org/10.1039/C8EE03369B} {\bibfield  {journal}
  {\bibinfo  {journal} {Energy Environ. Sci.}\ }\textbf {\bibinfo {volume}
  {12}},\ \bibinfo {pages} {1219} (\bibinfo {year} {2019})}\BibitemShut
  {NoStop}%
\bibitem [{\citenamefont {Ambrosio}\ \emph {et~al.}(2018)\citenamefont
  {Ambrosio}, \citenamefont {Wiktor}, \citenamefont {De~Angelis},\ and\
  \citenamefont {Pasquarello}}]{ambrosio2018origin}%
  \BibitemOpen
  \bibfield  {author} {\bibinfo {author} {\bibfnamefont {F.}~\bibnamefont
  {Ambrosio}}, \bibinfo {author} {\bibfnamefont {J.}~\bibnamefont {Wiktor}},
  \bibinfo {author} {\bibfnamefont {F.}~\bibnamefont {De~Angelis}}, \ and\
  \bibinfo {author} {\bibfnamefont {A.}~\bibnamefont {Pasquarello}},\ }\href
  {https://doi.org/10.1039/C7EE01981E} {\bibfield  {journal} {\bibinfo
  {journal} {Energy Environ. Sci.}\ }\textbf {\bibinfo {volume} {11}},\
  \bibinfo {pages} {101} (\bibinfo {year} {2018})}\BibitemShut {NoStop}%
\bibitem [{\citenamefont {Ambrosio}\ \emph {et~al.}(2019)\citenamefont
  {Ambrosio}, \citenamefont {Meggiolaro}, \citenamefont {Mosconi},\ and\
  \citenamefont {De~Angelis}}]{ambrosio2019charge}%
  \BibitemOpen
  \bibfield  {author} {\bibinfo {author} {\bibfnamefont {F.}~\bibnamefont
  {Ambrosio}}, \bibinfo {author} {\bibfnamefont {D.}~\bibnamefont
  {Meggiolaro}}, \bibinfo {author} {\bibfnamefont {E.}~\bibnamefont {Mosconi}},
  \ and\ \bibinfo {author} {\bibfnamefont {F.}~\bibnamefont {De~Angelis}},\
  }\href {https://doi.org/10.1021/acsenergylett.9b01353} {\bibfield  {journal}
  {\bibinfo  {journal} {ACS Energy Lett.}\ }\textbf {\bibinfo {volume} {4}},\
  \bibinfo {pages} {2013} (\bibinfo {year} {2019})}\BibitemShut {NoStop}%
\bibitem [{\citenamefont {Wang}\ \emph {et~al.}(2022)\citenamefont {Wang},
  \citenamefont {Chu}, \citenamefont {Huber}, \citenamefont {Tu}, \citenamefont
  {Dai}, \citenamefont {Wang}, \citenamefont {Peng}, \citenamefont {Zhao},\
  and\ \citenamefont {Zhu}}]{wang2022phonon}%
  \BibitemOpen
  \bibfield  {author} {\bibinfo {author} {\bibfnamefont {F.}~\bibnamefont
  {Wang}}, \bibinfo {author} {\bibfnamefont {W.}~\bibnamefont {Chu}}, \bibinfo
  {author} {\bibfnamefont {L.}~\bibnamefont {Huber}}, \bibinfo {author}
  {\bibfnamefont {T.}~\bibnamefont {Tu}}, \bibinfo {author} {\bibfnamefont
  {Y.}~\bibnamefont {Dai}}, \bibinfo {author} {\bibfnamefont {J.}~\bibnamefont
  {Wang}}, \bibinfo {author} {\bibfnamefont {H.}~\bibnamefont {Peng}}, \bibinfo
  {author} {\bibfnamefont {J.}~\bibnamefont {Zhao}}, \ and\ \bibinfo {author}
  {\bibfnamefont {X.-Y.}\ \bibnamefont {Zhu}},\ }\href
  {https://doi.org/10.1073/pnas.2122436119} {\bibfield  {journal} {\bibinfo
  {journal} {Proc. Natl. Acad. Sci. U.S.A.}\ }\textbf {\bibinfo {volume}
  {119}},\ \bibinfo {pages} {e2122436119} (\bibinfo {year} {2022})}\BibitemShut
  {NoStop}%
\bibitem [{\citenamefont {Miyata}\ \emph
  {et~al.}(2017{\natexlab{b}})\citenamefont {Miyata}, \citenamefont
  {Meggiolaro}, \citenamefont {Trinh}, \citenamefont {Joshi}, \citenamefont
  {Mosconi}, \citenamefont {Jones}, \citenamefont {De~Angelis},\ and\
  \citenamefont {Zhu}}]{miyata2017large}%
  \BibitemOpen
  \bibfield  {author} {\bibinfo {author} {\bibfnamefont {K.}~\bibnamefont
  {Miyata}}, \bibinfo {author} {\bibfnamefont {D.}~\bibnamefont {Meggiolaro}},
  \bibinfo {author} {\bibfnamefont {M.~T.}\ \bibnamefont {Trinh}}, \bibinfo
  {author} {\bibfnamefont {P.~P.}\ \bibnamefont {Joshi}}, \bibinfo {author}
  {\bibfnamefont {E.}~\bibnamefont {Mosconi}}, \bibinfo {author} {\bibfnamefont
  {S.~C.}\ \bibnamefont {Jones}}, \bibinfo {author} {\bibfnamefont
  {F.}~\bibnamefont {De~Angelis}}, \ and\ \bibinfo {author} {\bibfnamefont
  {X.-Y.}\ \bibnamefont {Zhu}},\ }\href
  {https://doi.org/10.1126/sciadv.1701217} {\bibfield  {journal} {\bibinfo
  {journal} {Sci. Adv.}\ }\textbf {\bibinfo {volume} {3}},\ \bibinfo {pages}
  {e1701217} (\bibinfo {year} {2017}{\natexlab{b}})}\BibitemShut {NoStop}%
\bibitem [{\citenamefont {Miyata}\ and\ \citenamefont
  {Zhu}(2018)}]{miyata2018ferroelectric}%
  \BibitemOpen
  \bibfield  {author} {\bibinfo {author} {\bibfnamefont {K.}~\bibnamefont
  {Miyata}}\ and\ \bibinfo {author} {\bibfnamefont {X.-Y.}\ \bibnamefont
  {Zhu}},\ }\href {https://doi.org/10.1038/s41563-018-0068-7} {\bibfield
  {journal} {\bibinfo  {journal} {Nat. Mater.}\ }\textbf {\bibinfo {volume}
  {17}},\ \bibinfo {pages} {379} (\bibinfo {year} {2018})}\BibitemShut
  {NoStop}%
\bibitem [{\citenamefont {Wang}\ \emph {et~al.}(2020)\citenamefont {Wang},
  \citenamefont {Fu}, \citenamefont {Ziffer}, \citenamefont {Dai},
  \citenamefont {Maehrlein},\ and\ \citenamefont {Zhu}}]{wang2020solvated}%
  \BibitemOpen
  \bibfield  {author} {\bibinfo {author} {\bibfnamefont {F.}~\bibnamefont
  {Wang}}, \bibinfo {author} {\bibfnamefont {Y.}~\bibnamefont {Fu}}, \bibinfo
  {author} {\bibfnamefont {M.~E.}\ \bibnamefont {Ziffer}}, \bibinfo {author}
  {\bibfnamefont {Y.}~\bibnamefont {Dai}}, \bibinfo {author} {\bibfnamefont
  {S.~F.}\ \bibnamefont {Maehrlein}}, \ and\ \bibinfo {author} {\bibfnamefont
  {X.-Y.}\ \bibnamefont {Zhu}},\ }\href {https://doi.org/10.1021/jacs.0c10943}
  {\bibfield  {journal} {\bibinfo  {journal} {J. Am. Chem. Soc.}\ }\textbf
  {\bibinfo {volume} {143}},\ \bibinfo {pages} {5} (\bibinfo {year}
  {2020})}\BibitemShut {NoStop}%
\bibitem [{\citenamefont {Fr{\" o}hlich}(1954)}]{Froehlich1954}%
  \BibitemOpen
  \bibfield  {author} {\bibinfo {author} {\bibfnamefont {H.}~\bibnamefont
  {Fr{\" o}hlich}},\ }\href {\doibase 10.1080/00018735400101213} {\bibfield
  {journal} {\bibinfo  {journal} {Adv. Phys.}\ }\textbf {\bibinfo {volume}
  {3}},\ \bibinfo {pages} {325} (\bibinfo {year} {1954})}\BibitemShut {NoStop}%
\bibitem [{\citenamefont {Alexandrov}\ and\ \citenamefont
  {Devreese}(2010)}]{AlexandrovDevreese2010}%
  \BibitemOpen
  \bibfield  {author} {\bibinfo {author} {\bibfnamefont {A.~S.}\ \bibnamefont
  {Alexandrov}}\ and\ \bibinfo {author} {\bibfnamefont {J.~T.}\ \bibnamefont
  {Devreese}},\ }\href {\doibase 10.1007/978-3-642-01896-1_1} {\emph {\bibinfo
  {title} {Advances in Polaron Physics}}}\ (\bibinfo  {publisher} {Springer},\
  \bibinfo {address} {Berlin Heidelberg,~},\ \bibinfo {year}
  {2010})\BibitemShut {NoStop}%
\bibitem [{\citenamefont {Zhu}\ \emph {et~al.}(2016)\citenamefont {Zhu},
  \citenamefont {Miyata}, \citenamefont {Fu}, \citenamefont {Wang},
  \citenamefont {Joshi}, \citenamefont {Niesner}, \citenamefont {Williams},
  \citenamefont {Jin},\ and\ \citenamefont {Zhu}}]{ZhuMiyata2016}%
  \BibitemOpen
  \bibfield  {author} {\bibinfo {author} {\bibfnamefont {H.}~\bibnamefont
  {Zhu}}, \bibinfo {author} {\bibfnamefont {K.}~\bibnamefont {Miyata}},
  \bibinfo {author} {\bibfnamefont {Y.}~\bibnamefont {Fu}}, \bibinfo {author}
  {\bibfnamefont {J.}~\bibnamefont {Wang}}, \bibinfo {author} {\bibfnamefont
  {P.~P.}\ \bibnamefont {Joshi}}, \bibinfo {author} {\bibfnamefont
  {D.}~\bibnamefont {Niesner}}, \bibinfo {author} {\bibfnamefont {K.~W.}\
  \bibnamefont {Williams}}, \bibinfo {author} {\bibfnamefont {S.}~\bibnamefont
  {Jin}}, \ and\ \bibinfo {author} {\bibfnamefont {X.-Y.}\ \bibnamefont
  {Zhu}},\ }\href {\doibase 10.1126/science.aaf9570} {\bibfield  {journal}
  {\bibinfo  {journal} {Science}\ }\textbf {\bibinfo {volume} {353}},\ \bibinfo
  {pages} {1409} (\bibinfo {year} {2016})}\BibitemShut {NoStop}%
\bibitem [{\citenamefont {Rakita}\ \emph {et~al.}(2017)\citenamefont {Rakita},
  \citenamefont {Bar-Elli}, \citenamefont {Meirzadeh}, \citenamefont {Kaslasi},
  \citenamefont {Peleg}, \citenamefont {Hodes}, \citenamefont {Lubomirsky},
  \citenamefont {Oron}, \citenamefont {Ehre},\ and\ \citenamefont
  {Cahen}}]{RakitaOmri2017}%
  \BibitemOpen
  \bibfield  {author} {\bibinfo {author} {\bibfnamefont {Y.}~\bibnamefont
  {Rakita}}, \bibinfo {author} {\bibfnamefont {O.}~\bibnamefont {Bar-Elli}},
  \bibinfo {author} {\bibfnamefont {E.}~\bibnamefont {Meirzadeh}}, \bibinfo
  {author} {\bibfnamefont {H.}~\bibnamefont {Kaslasi}}, \bibinfo {author}
  {\bibfnamefont {Y.}~\bibnamefont {Peleg}}, \bibinfo {author} {\bibfnamefont
  {G.}~\bibnamefont {Hodes}}, \bibinfo {author} {\bibfnamefont
  {I.}~\bibnamefont {Lubomirsky}}, \bibinfo {author} {\bibfnamefont
  {D.}~\bibnamefont {Oron}}, \bibinfo {author} {\bibfnamefont {D.}~\bibnamefont
  {Ehre}}, \ and\ \bibinfo {author} {\bibfnamefont {D.}~\bibnamefont {Cahen}},\
  }\href {\doibase 10.1073/pnas.1702429114} {\bibfield  {journal} {\bibinfo
  {journal} {Proc. Natl. Acad. Sci. U.S.A.}\ }\textbf {\bibinfo {volume}
  {114}},\ \bibinfo {pages} {E5504} (\bibinfo {year} {2017})}\BibitemShut
  {NoStop}%
\bibitem [{\citenamefont {Shahrokhi}\ \emph {et~al.}(2020)\citenamefont
  {Shahrokhi}, \citenamefont {Gao}, \citenamefont {Wang}, \citenamefont
  {Anandan}, \citenamefont {Rahaman}, \citenamefont {Singh}, \citenamefont
  {Wang}, \citenamefont {Cazorla}, \citenamefont {Yuan}, \citenamefont {Liu},\
  and\ \citenamefont {Wu}}]{ShahrokhiGao2020}%
  \BibitemOpen
  \bibfield  {author} {\bibinfo {author} {\bibfnamefont {S.}~\bibnamefont
  {Shahrokhi}}, \bibinfo {author} {\bibfnamefont {W.}~\bibnamefont {Gao}},
  \bibinfo {author} {\bibfnamefont {Y.}~\bibnamefont {Wang}}, \bibinfo {author}
  {\bibfnamefont {P.~R.}\ \bibnamefont {Anandan}}, \bibinfo {author}
  {\bibfnamefont {M.~Z.}\ \bibnamefont {Rahaman}}, \bibinfo {author}
  {\bibfnamefont {S.}~\bibnamefont {Singh}}, \bibinfo {author} {\bibfnamefont
  {D.}~\bibnamefont {Wang}}, \bibinfo {author} {\bibfnamefont {C.}~\bibnamefont
  {Cazorla}}, \bibinfo {author} {\bibfnamefont {G.}~\bibnamefont {Yuan}},
  \bibinfo {author} {\bibfnamefont {J.-M.}\ \bibnamefont {Liu}}, \ and\
  \bibinfo {author} {\bibfnamefont {T.}~\bibnamefont {Wu}},\ }\href {\doibase
  https://doi.org/10.1002/smtd.202000149} {\bibfield  {journal} {\bibinfo
  {journal} {Small Methods}\ }\textbf {\bibinfo {volume} {4}},\ \bibinfo
  {pages} {2000149} (\bibinfo {year} {2020})}\BibitemShut {NoStop}%
\bibitem [{\citenamefont {Liu}\ \emph {et~al.}(2018)\citenamefont {Liu},
  \citenamefont {Collins}, \citenamefont {Proksch}, \citenamefont {Kim},
  \citenamefont {Watson}, \citenamefont {Doughty}, \citenamefont {Calhoun},
  \citenamefont {Ahmadi}, \citenamefont {Ievlev}, \citenamefont {Jesse},
  \citenamefont {Retterer}, \citenamefont {Belianinov}, \citenamefont {Xiao},
  \citenamefont {Huang}, \citenamefont {Sumpter}, \citenamefont {Kalinin},
  \citenamefont {Hu},\ and\ \citenamefont {Ovchinnikova}}]{LiuCollins2018}%
  \BibitemOpen
  \bibfield  {author} {\bibinfo {author} {\bibfnamefont {Y.}~\bibnamefont
  {Liu}}, \bibinfo {author} {\bibfnamefont {L.}~\bibnamefont {Collins}},
  \bibinfo {author} {\bibfnamefont {R.}~\bibnamefont {Proksch}}, \bibinfo
  {author} {\bibfnamefont {S.}~\bibnamefont {Kim}}, \bibinfo {author}
  {\bibfnamefont {B.~R.}\ \bibnamefont {Watson}}, \bibinfo {author}
  {\bibfnamefont {B.}~\bibnamefont {Doughty}}, \bibinfo {author} {\bibfnamefont
  {T.~R.}\ \bibnamefont {Calhoun}}, \bibinfo {author} {\bibfnamefont
  {M.}~\bibnamefont {Ahmadi}}, \bibinfo {author} {\bibfnamefont {A.~V.}\
  \bibnamefont {Ievlev}}, \bibinfo {author} {\bibfnamefont {S.}~\bibnamefont
  {Jesse}}, \bibinfo {author} {\bibfnamefont {S.~T.}\ \bibnamefont {Retterer}},
  \bibinfo {author} {\bibfnamefont {A.}~\bibnamefont {Belianinov}}, \bibinfo
  {author} {\bibfnamefont {K.}~\bibnamefont {Xiao}}, \bibinfo {author}
  {\bibfnamefont {J.}~\bibnamefont {Huang}}, \bibinfo {author} {\bibfnamefont
  {B.~G.}\ \bibnamefont {Sumpter}}, \bibinfo {author} {\bibfnamefont {S.~V.}\
  \bibnamefont {Kalinin}}, \bibinfo {author} {\bibfnamefont {B.}~\bibnamefont
  {Hu}}, \ and\ \bibinfo {author} {\bibfnamefont {O.~S.}\ \bibnamefont
  {Ovchinnikova}},\ }\href {\doibase 10.1038/s41563-018-0152-z} {\bibfield
  {journal} {\bibinfo  {journal} {Nat. Mater.}\ }\textbf {\bibinfo {volume}
  {17}},\ \bibinfo {pages} {1013} (\bibinfo {year} {2018})}\BibitemShut
  {NoStop}%
\bibitem [{\citenamefont {Alexandrov}\ and\ \citenamefont
  {Yavidov}(2004)}]{AlexandrovYavidov2004}%
  \BibitemOpen
  \bibfield  {author} {\bibinfo {author} {\bibfnamefont {A.~S.}\ \bibnamefont
  {Alexandrov}}\ and\ \bibinfo {author} {\bibfnamefont {B.~Y.}\ \bibnamefont
  {Yavidov}},\ }\href {\doibase 10.1103/PhysRevB.69.073101} {\bibfield
  {journal} {\bibinfo  {journal} {Phys. Rev. B}\ }\textbf {\bibinfo {volume}
  {69}},\ \bibinfo {pages} {073101} (\bibinfo {year} {2004})}\BibitemShut
  {NoStop}%
\bibitem [{\citenamefont {Tozer}\ and\ \citenamefont
  {Barford}(2014)}]{TozerBarfold2014}%
  \BibitemOpen
  \bibfield  {author} {\bibinfo {author} {\bibfnamefont {O.~R.}\ \bibnamefont
  {Tozer}}\ and\ \bibinfo {author} {\bibfnamefont {W.}~\bibnamefont
  {Barford}},\ }\href {\doibase 10.1103/PhysRevB.89.155434} {\bibfield
  {journal} {\bibinfo  {journal} {Phys. Rev. B}\ }\textbf {\bibinfo {volume}
  {89}},\ \bibinfo {pages} {155434} (\bibinfo {year} {2014})}\BibitemShut
  {NoStop}%
\bibitem [{\citenamefont {Tailor}\ and\ \citenamefont
  {Satapathi}(2022)}]{TailorSatapathi2022}%
  \BibitemOpen
  \bibfield  {author} {\bibinfo {author} {\bibfnamefont {N.~K.}\ \bibnamefont
  {Tailor}}\ and\ \bibinfo {author} {\bibfnamefont {S.}~\bibnamefont
  {Satapathi}},\ }\href {\doibase 10.1021/acs.jpcc.2c05534} {\bibfield
  {journal} {\bibinfo  {journal} {J. Phys. Chem. C}\ }\textbf {\bibinfo
  {volume} {126}},\ \bibinfo {pages} {17789} (\bibinfo {year}
  {2022})}\BibitemShut {NoStop}%
\bibitem [{\citenamefont {Baimuratov}\ \emph {et~al.}(2017)\citenamefont
  {Baimuratov}, \citenamefont {Pereziabova}, \citenamefont {Zhu}, \citenamefont
  {Leonov}, \citenamefont {Baranov}, \citenamefont {Fedorov},\ and\
  \citenamefont {Rukhlenko}}]{BaimuratovPereziabova2017}%
  \BibitemOpen
  \bibfield  {author} {\bibinfo {author} {\bibfnamefont {A.~S.}\ \bibnamefont
  {Baimuratov}}, \bibinfo {author} {\bibfnamefont {T.~P.}\ \bibnamefont
  {Pereziabova}}, \bibinfo {author} {\bibfnamefont {W.}~\bibnamefont {Zhu}},
  \bibinfo {author} {\bibfnamefont {M.~Y.}\ \bibnamefont {Leonov}}, \bibinfo
  {author} {\bibfnamefont {A.~V.}\ \bibnamefont {Baranov}}, \bibinfo {author}
  {\bibfnamefont {A.~V.}\ \bibnamefont {Fedorov}}, \ and\ \bibinfo {author}
  {\bibfnamefont {I.~D.}\ \bibnamefont {Rukhlenko}},\ }\href {\doibase
  10.1021/acs.nanolett.7b02203} {\bibfield  {journal} {\bibinfo  {journal}
  {Nano Lett.}\ }\textbf {\bibinfo {volume} {17}},\ \bibinfo {pages} {5514}
  (\bibinfo {year} {2017})}\BibitemShut {NoStop}%
\bibitem [{\citenamefont {Jiao}\ \emph {et~al.}(2021)\citenamefont {Jiao},
  \citenamefont {Yi}, \citenamefont {Wang}, \citenamefont {Li}, \citenamefont
  {Hao}, \citenamefont {Pan}, \citenamefont {Shi}, \citenamefont {Li},
  \citenamefont {Liu}, \citenamefont {Zhang}, \citenamefont {Gao},
  \citenamefont {Zhao},\ and\ \citenamefont {Lu}}]{JiaoYi2021}%
  \BibitemOpen
  \bibfield  {author} {\bibinfo {author} {\bibfnamefont {Y.}~\bibnamefont
  {Jiao}}, \bibinfo {author} {\bibfnamefont {S.}~\bibnamefont {Yi}}, \bibinfo
  {author} {\bibfnamefont {H.}~\bibnamefont {Wang}}, \bibinfo {author}
  {\bibfnamefont {B.}~\bibnamefont {Li}}, \bibinfo {author} {\bibfnamefont
  {W.}~\bibnamefont {Hao}}, \bibinfo {author} {\bibfnamefont {L.}~\bibnamefont
  {Pan}}, \bibinfo {author} {\bibfnamefont {Y.}~\bibnamefont {Shi}}, \bibinfo
  {author} {\bibfnamefont {X.}~\bibnamefont {Li}}, \bibinfo {author}
  {\bibfnamefont {P.}~\bibnamefont {Liu}}, \bibinfo {author} {\bibfnamefont
  {H.}~\bibnamefont {Zhang}}, \bibinfo {author} {\bibfnamefont
  {C.}~\bibnamefont {Gao}}, \bibinfo {author} {\bibfnamefont {J.}~\bibnamefont
  {Zhao}}, \ and\ \bibinfo {author} {\bibfnamefont {J.}~\bibnamefont {Lu}},\
  }\href {\doibase https://doi.org/10.1002/adfm.202006243} {\bibfield
  {journal} {\bibinfo  {journal} {Adv. Func. Mater.}\ }\textbf {\bibinfo
  {volume} {31}},\ \bibinfo {pages} {2006243} (\bibinfo {year}
  {2021})}\BibitemShut {NoStop}%
\bibitem [{\citenamefont {Fabini}\ \emph {et~al.}(2017)\citenamefont {Fabini},
  \citenamefont {Siaw}, \citenamefont {Stoumpos}, \citenamefont {Laurita},
  \citenamefont {Olds}, \citenamefont {Page}, \citenamefont {Hu}, \citenamefont
  {Kanatzidis}, \citenamefont {Han},\ and\ \citenamefont
  {Seshadri}}]{FabiniSiaw2017}%
  \BibitemOpen
  \bibfield  {author} {\bibinfo {author} {\bibfnamefont {D.~H.}\ \bibnamefont
  {Fabini}}, \bibinfo {author} {\bibfnamefont {T.~A.}\ \bibnamefont {Siaw}},
  \bibinfo {author} {\bibfnamefont {C.~C.}\ \bibnamefont {Stoumpos}}, \bibinfo
  {author} {\bibfnamefont {G.}~\bibnamefont {Laurita}}, \bibinfo {author}
  {\bibfnamefont {D.}~\bibnamefont {Olds}}, \bibinfo {author} {\bibfnamefont
  {K.}~\bibnamefont {Page}}, \bibinfo {author} {\bibfnamefont {J.~G.}\
  \bibnamefont {Hu}}, \bibinfo {author} {\bibfnamefont {M.~G.}\ \bibnamefont
  {Kanatzidis}}, \bibinfo {author} {\bibfnamefont {S.}~\bibnamefont {Han}}, \
  and\ \bibinfo {author} {\bibfnamefont {R.}~\bibnamefont {Seshadri}},\ }\href
  {\doibase 10.1021/jacs.7b09536} {\bibfield  {journal} {\bibinfo  {journal}
  {J. Am. Chem. Soc.}\ }\textbf {\bibinfo {volume} {139}},\ \bibinfo {pages}
  {16875} (\bibinfo {year} {2017})}\BibitemShut {NoStop}%
\bibitem [{\citenamefont {Kang}\ and\ \citenamefont
  {Wang}(2017)}]{Kang2017dynamic}%
  \BibitemOpen
  \bibfield  {author} {\bibinfo {author} {\bibfnamefont {J.}~\bibnamefont
  {Kang}}\ and\ \bibinfo {author} {\bibfnamefont {L.-W.}\ \bibnamefont
  {Wang}},\ }\href {\doibase 10.1021/acs.jpclett.7b01501} {\bibfield  {journal}
  {\bibinfo  {journal} {J. Phys. Chem. Lett.}\ }\textbf {\bibinfo {volume}
  {8}},\ \bibinfo {pages} {3875} (\bibinfo {year} {2017})}\BibitemShut
  {NoStop}%
\bibitem [{\citenamefont {Lang}\ and\ \citenamefont
  {Firsov}(1963)}]{LangFirsov1963}%
  \BibitemOpen
  \bibfield  {author} {\bibinfo {author} {\bibfnamefont {I.~G.}\ \bibnamefont
  {Lang}}\ and\ \bibinfo {author} {\bibfnamefont {Y.~A.}\ \bibnamefont
  {Firsov}},\ }\href
  {http://www.jetp.ras.ru/cgi-bin/e/index/r/45/1/p378?a=list} {\bibfield
  {journal} {\bibinfo  {journal} {Zh. Eksp. Teor. Fiz.}\ }\textbf {\bibinfo
  {volume} {45}},\ \bibinfo {pages} {378} (\bibinfo {year} {1963})},\ \bibinfo
  {note} {[Sov. Phys. JETP 18, 262 (1964)]}\BibitemShut {NoStop}%
\bibitem [{\citenamefont {Grusdt}\ and\ \citenamefont
  {Demler}(2015)}]{grusdt2015new}%
  \BibitemOpen
  \bibfield  {author} {\bibinfo {author} {\bibfnamefont {F.}~\bibnamefont
  {Grusdt}}\ and\ \bibinfo {author} {\bibfnamefont {E.}~\bibnamefont
  {Demler}},\ }\href@noop {} {\bibfield  {journal} {\bibinfo  {journal}
  {Quantum Matter at Ultralow Temperatures}\ }\textbf {\bibinfo {volume}
  {191}},\ \bibinfo {pages} {325} (\bibinfo {year} {2015})}\BibitemShut
  {NoStop}%
\bibitem [{Note1()}]{Note1}%
  \BibitemOpen
  \bibinfo {note} {See Supplemental Material for details.}\BibitemShut {Stop}%
\bibitem [{\citenamefont {Spohn}(1986)}]{Spohn1986}%
  \BibitemOpen
  \bibfield  {author} {\bibinfo {author} {\bibfnamefont {H.}~\bibnamefont
  {Spohn}},\ }\href {\doibase 10.1088/0305-4470/19/4/014} {\bibfield  {journal}
  {\bibinfo  {journal} {J. Phys. A}\ }\textbf {\bibinfo {volume} {19}},\
  \bibinfo {pages} {533} (\bibinfo {year} {1986})}\BibitemShut {NoStop}%
\bibitem [{\citenamefont {Gerlach}\ and\ \citenamefont
  {L\"owen}(1991)}]{GerlachLowen1991}%
  \BibitemOpen
  \bibfield  {author} {\bibinfo {author} {\bibfnamefont {B.}~\bibnamefont
  {Gerlach}}\ and\ \bibinfo {author} {\bibfnamefont {H.}~\bibnamefont
  {L\"owen}},\ }\href {\doibase 10.1103/RevModPhys.63.63} {\bibfield  {journal}
  {\bibinfo  {journal} {Rev. Mod. Phys.}\ }\textbf {\bibinfo {volume} {63}},\
  \bibinfo {pages} {63} (\bibinfo {year} {1991})}\BibitemShut {NoStop}%
\bibitem [{\citenamefont {Holstein}(1959{\natexlab{a}})}]{Holstein1959a}%
  \BibitemOpen
  \bibfield  {author} {\bibinfo {author} {\bibfnamefont {T.}~\bibnamefont
  {Holstein}},\ }\href {\doibase https://doi.org/10.1016/0003-4916(59)90002-8}
  {\bibfield  {journal} {\bibinfo  {journal} {Ann. Phys.}\ }\textbf {\bibinfo
  {volume} {8}},\ \bibinfo {pages} {325 } (\bibinfo {year}
  {1959}{\natexlab{a}})}\BibitemShut {NoStop}%
\bibitem [{\citenamefont {Holstein}(1959{\natexlab{b}})}]{Holstein1959b}%
  \BibitemOpen
  \bibfield  {author} {\bibinfo {author} {\bibfnamefont {T.}~\bibnamefont
  {Holstein}},\ }\href {\doibase https://doi.org/10.1016/0003-4916(59)90003-X}
  {\bibfield  {journal} {\bibinfo  {journal} {Ann. Phys.}\ }\textbf {\bibinfo
  {volume} {8}},\ \bibinfo {pages} {343 } (\bibinfo {year}
  {1959}{\natexlab{b}})}\BibitemShut {NoStop}%
\bibitem [{\citenamefont {Stranks}\ \emph {et~al.}(2013)\citenamefont
  {Stranks}, \citenamefont {Eperon}, \citenamefont {Grancini}, \citenamefont
  {Menelaou}, \citenamefont {Alcocer}, \citenamefont {Leijtens}, \citenamefont
  {Herz}, \citenamefont {Petrozza},\ and\ \citenamefont
  {Snaith}}]{StranksEperon2013}%
  \BibitemOpen
  \bibfield  {author} {\bibinfo {author} {\bibfnamefont {S.~D.}\ \bibnamefont
  {Stranks}}, \bibinfo {author} {\bibfnamefont {G.~E.}\ \bibnamefont {Eperon}},
  \bibinfo {author} {\bibfnamefont {G.}~\bibnamefont {Grancini}}, \bibinfo
  {author} {\bibfnamefont {C.}~\bibnamefont {Menelaou}}, \bibinfo {author}
  {\bibfnamefont {M.~J.~P.}\ \bibnamefont {Alcocer}}, \bibinfo {author}
  {\bibfnamefont {T.}~\bibnamefont {Leijtens}}, \bibinfo {author}
  {\bibfnamefont {L.~M.}\ \bibnamefont {Herz}}, \bibinfo {author}
  {\bibfnamefont {A.}~\bibnamefont {Petrozza}}, \ and\ \bibinfo {author}
  {\bibfnamefont {H.~J.}\ \bibnamefont {Snaith}},\ }\href {\doibase
  10.1126/science.1243982} {\bibfield  {journal} {\bibinfo  {journal}
  {Science}\ }\textbf {\bibinfo {volume} {342}},\ \bibinfo {pages} {341}
  (\bibinfo {year} {2013})}\BibitemShut {NoStop}%
\bibitem [{\citenamefont {Shi}\ \emph {et~al.}(2015)\citenamefont {Shi},
  \citenamefont {Adinolfi}, \citenamefont {Comin}, \citenamefont {Yuan},
  \citenamefont {Alarousu}, \citenamefont {Buin}, \citenamefont {Chen},
  \citenamefont {Hoogland}, \citenamefont {Rothenberger}, \citenamefont
  {Katsiev}, \citenamefont {Losovyj}, \citenamefont {Zhang}, \citenamefont
  {Dowben}, \citenamefont {Mohammed}, \citenamefont {Sargent},\ and\
  \citenamefont {Bakr}}]{ShiAdinolfi2015}%
  \BibitemOpen
  \bibfield  {author} {\bibinfo {author} {\bibfnamefont {D.}~\bibnamefont
  {Shi}}, \bibinfo {author} {\bibfnamefont {V.}~\bibnamefont {Adinolfi}},
  \bibinfo {author} {\bibfnamefont {R.}~\bibnamefont {Comin}}, \bibinfo
  {author} {\bibfnamefont {M.}~\bibnamefont {Yuan}}, \bibinfo {author}
  {\bibfnamefont {E.}~\bibnamefont {Alarousu}}, \bibinfo {author}
  {\bibfnamefont {A.}~\bibnamefont {Buin}}, \bibinfo {author} {\bibfnamefont
  {Y.}~\bibnamefont {Chen}}, \bibinfo {author} {\bibfnamefont {S.}~\bibnamefont
  {Hoogland}}, \bibinfo {author} {\bibfnamefont {A.}~\bibnamefont
  {Rothenberger}}, \bibinfo {author} {\bibfnamefont {K.}~\bibnamefont
  {Katsiev}}, \bibinfo {author} {\bibfnamefont {Y.}~\bibnamefont {Losovyj}},
  \bibinfo {author} {\bibfnamefont {X.}~\bibnamefont {Zhang}}, \bibinfo
  {author} {\bibfnamefont {P.~A.}\ \bibnamefont {Dowben}}, \bibinfo {author}
  {\bibfnamefont {O.~F.}\ \bibnamefont {Mohammed}}, \bibinfo {author}
  {\bibfnamefont {E.~H.}\ \bibnamefont {Sargent}}, \ and\ \bibinfo {author}
  {\bibfnamefont {O.~M.}\ \bibnamefont {Bakr}},\ }\href {\doibase
  10.1126/science.aaa2725} {\bibfield  {journal} {\bibinfo  {journal}
  {Science}\ }\textbf {\bibinfo {volume} {347}},\ \bibinfo {pages} {519}
  (\bibinfo {year} {2015})}\BibitemShut {NoStop}%
\bibitem [{\citenamefont {Dong}\ \emph {et~al.}(2015)\citenamefont {Dong},
  \citenamefont {Fang}, \citenamefont {Shao}, \citenamefont {Mulligan},
  \citenamefont {Qiu}, \citenamefont {Cao},\ and\ \citenamefont
  {Huang}}]{DongFang2015}%
  \BibitemOpen
  \bibfield  {author} {\bibinfo {author} {\bibfnamefont {Q.}~\bibnamefont
  {Dong}}, \bibinfo {author} {\bibfnamefont {Y.}~\bibnamefont {Fang}}, \bibinfo
  {author} {\bibfnamefont {Y.}~\bibnamefont {Shao}}, \bibinfo {author}
  {\bibfnamefont {P.}~\bibnamefont {Mulligan}}, \bibinfo {author}
  {\bibfnamefont {J.}~\bibnamefont {Qiu}}, \bibinfo {author} {\bibfnamefont
  {L.}~\bibnamefont {Cao}}, \ and\ \bibinfo {author} {\bibfnamefont
  {J.}~\bibnamefont {Huang}},\ }\href {\doibase 10.1126/science.aaa5760}
  {\bibfield  {journal} {\bibinfo  {journal} {Science}\ }\textbf {\bibinfo
  {volume} {347}},\ \bibinfo {pages} {967} (\bibinfo {year}
  {2015})}\BibitemShut {NoStop}%
\bibitem [{\citenamefont {Price}\ \emph {et~al.}(2015)\citenamefont {Price},
  \citenamefont {Butkus}, \citenamefont {Jellicoe}, \citenamefont {Sadhanala},
  \citenamefont {Briane}, \citenamefont {Halpert}, \citenamefont {Broch},
  \citenamefont {Hodgkiss}, \citenamefont {Friend},\ and\ \citenamefont
  {Deschler}}]{Price2015}%
  \BibitemOpen
  \bibfield  {author} {\bibinfo {author} {\bibfnamefont {M.~B.}\ \bibnamefont
  {Price}}, \bibinfo {author} {\bibfnamefont {J.}~\bibnamefont {Butkus}},
  \bibinfo {author} {\bibfnamefont {T.~C.}\ \bibnamefont {Jellicoe}}, \bibinfo
  {author} {\bibfnamefont {A.}~\bibnamefont {Sadhanala}}, \bibinfo {author}
  {\bibfnamefont {A.}~\bibnamefont {Briane}}, \bibinfo {author} {\bibfnamefont
  {J.~E.}\ \bibnamefont {Halpert}}, \bibinfo {author} {\bibfnamefont
  {K.}~\bibnamefont {Broch}}, \bibinfo {author} {\bibfnamefont {J.~M.}\
  \bibnamefont {Hodgkiss}}, \bibinfo {author} {\bibfnamefont {R.~H.}\
  \bibnamefont {Friend}}, \ and\ \bibinfo {author} {\bibfnamefont
  {F.}~\bibnamefont {Deschler}},\ }\href {\doibase 10.1038/ncomms9420}
  {\bibfield  {journal} {\bibinfo  {journal} {Nat. Commun.}\ }\textbf {\bibinfo
  {volume} {6}},\ \bibinfo {pages} {8420} (\bibinfo {year} {2015})}\BibitemShut
  {NoStop}%
\bibitem [{\citenamefont {Jeckelmann}\ and\ \citenamefont
  {White}(1998)}]{JackelmannWhite1998}%
  \BibitemOpen
  \bibfield  {author} {\bibinfo {author} {\bibfnamefont {E.}~\bibnamefont
  {Jeckelmann}}\ and\ \bibinfo {author} {\bibfnamefont {S.~R.}\ \bibnamefont
  {White}},\ }\href {\doibase 10.1103/PhysRevB.57.6376} {\bibfield  {journal}
  {\bibinfo  {journal} {Phys. Rev. B}\ }\textbf {\bibinfo {volume} {57}},\
  \bibinfo {pages} {6376} (\bibinfo {year} {1998})}\BibitemShut {NoStop}%
\bibitem [{\citenamefont {Schmidt}\ and\ \citenamefont
  {Lemeshko}(2015)}]{schmidt2015rotation}%
  \BibitemOpen
  \bibfield  {author} {\bibinfo {author} {\bibfnamefont {R.}~\bibnamefont
  {Schmidt}}\ and\ \bibinfo {author} {\bibfnamefont {M.}~\bibnamefont
  {Lemeshko}},\ }\href {https://doi.org/10.1103/PhysRevLett.114.203001}
  {\bibfield  {journal} {\bibinfo  {journal} {Phys. Rev. Lett.}\ }\textbf
  {\bibinfo {volume} {114}},\ \bibinfo {pages} {203001} (\bibinfo {year}
  {2015})}\BibitemShut {NoStop}%
\bibitem [{\citenamefont {Schmidt}\ and\ \citenamefont
  {Lemeshko}(2016)}]{schmidt2016deformation}%
  \BibitemOpen
  \bibfield  {author} {\bibinfo {author} {\bibfnamefont {R.}~\bibnamefont
  {Schmidt}}\ and\ \bibinfo {author} {\bibfnamefont {M.}~\bibnamefont
  {Lemeshko}},\ }\href {https://doi.org/10.1103/PhysRevX.6.011012} {\bibfield
  {journal} {\bibinfo  {journal} {Phys. Rev. X}\ }\textbf {\bibinfo {volume}
  {6}},\ \bibinfo {pages} {011012} (\bibinfo {year} {2016})}\BibitemShut
  {NoStop}%
\bibitem [{\citenamefont {Yakaboylu}\ and\ \citenamefont
  {Lemeshko}(2017)}]{YakaboyluScrPRL17}%
  \BibitemOpen
  \bibfield  {author} {\bibinfo {author} {\bibfnamefont {E.}~\bibnamefont
  {Yakaboylu}}\ and\ \bibinfo {author} {\bibfnamefont {M.}~\bibnamefont
  {Lemeshko}},\ }\href {\doibase 10.1103/PhysRevLett.118.085302} {\bibfield
  {journal} {\bibinfo  {journal} {Phys. Rev. Lett.}\ }\textbf {\bibinfo
  {volume} {118}},\ \bibinfo {pages} {085302} (\bibinfo {year}
  {2017})}\BibitemShut {NoStop}%
\bibitem [{\citenamefont {Cui}\ \emph {et~al.}(2022)\citenamefont {Cui},
  \citenamefont {Liu}, \citenamefont {Deng}, \citenamefont {Zhang},
  \citenamefont {Yang}, \citenamefont {Li},\ and\ \citenamefont
  {Wang}}]{cui2022microscopic}%
  \BibitemOpen
  \bibfield  {author} {\bibinfo {author} {\bibfnamefont {Y.}~\bibnamefont
  {Cui}}, \bibinfo {author} {\bibfnamefont {Y.-Y.}\ \bibnamefont {Liu}},
  \bibinfo {author} {\bibfnamefont {J.-P.}\ \bibnamefont {Deng}}, \bibinfo
  {author} {\bibfnamefont {X.-Z.}\ \bibnamefont {Zhang}}, \bibinfo {author}
  {\bibfnamefont {R.-B.}\ \bibnamefont {Yang}}, \bibinfo {author}
  {\bibfnamefont {Z.-Q.}\ \bibnamefont {Li}}, \ and\ \bibinfo {author}
  {\bibfnamefont {Z.-W.}\ \bibnamefont {Wang}},\ }\href
  {https://doi.org/10.48550/arXiv.2209.13861} {\bibfield  {journal} {\bibinfo
  {journal} {arXiv:2209.13861}\ } (\bibinfo {year} {2022})}\BibitemShut
  {NoStop}%
\bibitem [{\citenamefont {Wu}\ \emph {et~al.}(2022)\citenamefont {Wu},
  \citenamefont {Cui}, \citenamefont {Li},\ and\ \citenamefont
  {Wang}}]{wu2022optical}%
  \BibitemOpen
  \bibfield  {author} {\bibinfo {author} {\bibfnamefont {J.-W.}\ \bibnamefont
  {Wu}}, \bibinfo {author} {\bibfnamefont {Y.}~\bibnamefont {Cui}}, \bibinfo
  {author} {\bibfnamefont {S.-J.}\ \bibnamefont {Li}}, \ and\ \bibinfo {author}
  {\bibfnamefont {Z.-W.}\ \bibnamefont {Wang}},\ }\href
  {https://doi.org/10.48550/arXiv.2212.06356} {\bibfield  {journal} {\bibinfo
  {journal} {arXiv:2212.06356}\ } (\bibinfo {year} {2022})}\BibitemShut
  {NoStop}%
\bibitem [{\citenamefont {Volosniev}\ \emph
  {et~al.}(2023{\natexlab{a}})\citenamefont {Volosniev}, \citenamefont {Kumar},
  \citenamefont {Lorenc}, \citenamefont {Ashourishokri}, \citenamefont
  {Zhumekenov}, \citenamefont {Bakr}, \citenamefont {Lemeshko},\ and\
  \citenamefont {Alpichshev}}]{Volosniev22a}%
  \BibitemOpen
  \bibfield  {author} {\bibinfo {author} {\bibfnamefont {A.~G.}\ \bibnamefont
  {Volosniev}}, \bibinfo {author} {\bibfnamefont {A.~S.}\ \bibnamefont
  {Kumar}}, \bibinfo {author} {\bibfnamefont {D.}~\bibnamefont {Lorenc}},
  \bibinfo {author} {\bibfnamefont {Y.}~\bibnamefont {Ashourishokri}}, \bibinfo
  {author} {\bibfnamefont {A.~A.}\ \bibnamefont {Zhumekenov}}, \bibinfo
  {author} {\bibfnamefont {O.~M.}\ \bibnamefont {Bakr}}, \bibinfo {author}
  {\bibfnamefont {M.}~\bibnamefont {Lemeshko}}, \ and\ \bibinfo {author}
  {\bibfnamefont {Z.}~\bibnamefont {Alpichshev}},\ }\href
  {https://arxiv.org/abs/2203.09443} {\bibfield  {journal} {\bibinfo  {journal}
  {Phys. Rev. Lett., in press, arXiv:2203.09443}\ } (\bibinfo {year}
  {2023}{\natexlab{a}})}\BibitemShut {NoStop}%
\bibitem [{\citenamefont {Volosniev}\ \emph
  {et~al.}(2023{\natexlab{b}})\citenamefont {Volosniev}, \citenamefont {Kumar},
  \citenamefont {Lorenc}, \citenamefont {Ashourishokri}, \citenamefont
  {Zhumekenov}, \citenamefont {Bakr}, \citenamefont {Lemeshko},\ and\
  \citenamefont {Alpichshev}}]{Volosniev22b}%
  \BibitemOpen
  \bibfield  {author} {\bibinfo {author} {\bibfnamefont {A.~G.}\ \bibnamefont
  {Volosniev}}, \bibinfo {author} {\bibfnamefont {A.~S.}\ \bibnamefont
  {Kumar}}, \bibinfo {author} {\bibfnamefont {D.}~\bibnamefont {Lorenc}},
  \bibinfo {author} {\bibfnamefont {Y.}~\bibnamefont {Ashourishokri}}, \bibinfo
  {author} {\bibfnamefont {A.~A.}\ \bibnamefont {Zhumekenov}}, \bibinfo
  {author} {\bibfnamefont {O.~M.}\ \bibnamefont {Bakr}}, \bibinfo {author}
  {\bibfnamefont {M.}~\bibnamefont {Lemeshko}}, \ and\ \bibinfo {author}
  {\bibfnamefont {Z.}~\bibnamefont {Alpichshev}},\ }\href
  {https://arxiv.org/abs/2204.04022} {\bibfield  {journal} {\bibinfo  {journal}
  {Phys. Rev. B., in press, arXiv:2204.04022}\ } (\bibinfo {year}
  {2023}{\natexlab{b}})}\BibitemShut {NoStop}%
\end{thebibliography}%

%%%%%%%%%% Merge with supplemental materials %%%%%%%%%%
%\pagebreak
\newpage~\\
%\newpage
% \widetext
\onecolumngrid
\begin{center}
    \textbf{\large Supplementary Material\\[3pt] Rotor Lattice Model of Ferroelectric Large Polarons}
\end{center}
\twocolumngrid
%%%%%%%%%% Merge with supplemental materials %%%%%%%%%%
%%%%%%%%%% Prefix a `\Sigma`` to all equations, figures, tables and reset the counter %%%%%%%%%%
\setcounter{equation}{0}
\setcounter{figure}{0}
\setcounter{table}{0}
\setcounter{page}{1}
\makeatletter
\renewcommand{\theequation}{S\arabic{equation}}
\renewcommand{\thefigure}{S\arabic{figure}}
\renewcommand{\bibnumfmt}[1]{[S#1]}
\renewcommand{\citenumfont}[1]{S#1} 
%%%%%%%%%% Prefix a ''S`` to all equations, figures, tables and reset the counter %%%%%%%%%%

%\renewcommand\thefigure{S\arabic{figure}}    
%\renewcommand\theequation{S\arabic{equation}}    
%\setcounter{figure}{0}   
%
\section{Lang-Firsov transformation}

The Lang-Firsov transformation has been successfully used to  describe polarons in the Holstein model \citeS{AlexandrovDevreese2010s, LangFirsov1963s, Holstein1959as, Holstein1959bs}. Within the Lang-Firsov transformation one diagonalizes the Hamiltonian, \(\hat{H}_{V_0 \to \infty}\) for \(t=0\), and then uses the unitary matrix \(\hat{U}\) obtained by the diagonalization to obtain the hopping term in the transformed frame. Since in the case of the Holstein polaron the phonon--electron interaction has the simple form of a potential gradient, the operator \(\hat{U}\) corresponds to a displacement operator for the phonons and therefore the transformation can be performed   analytically.

\begin{figure}[b] \centering \includegraphics[width=1.0\linewidth]{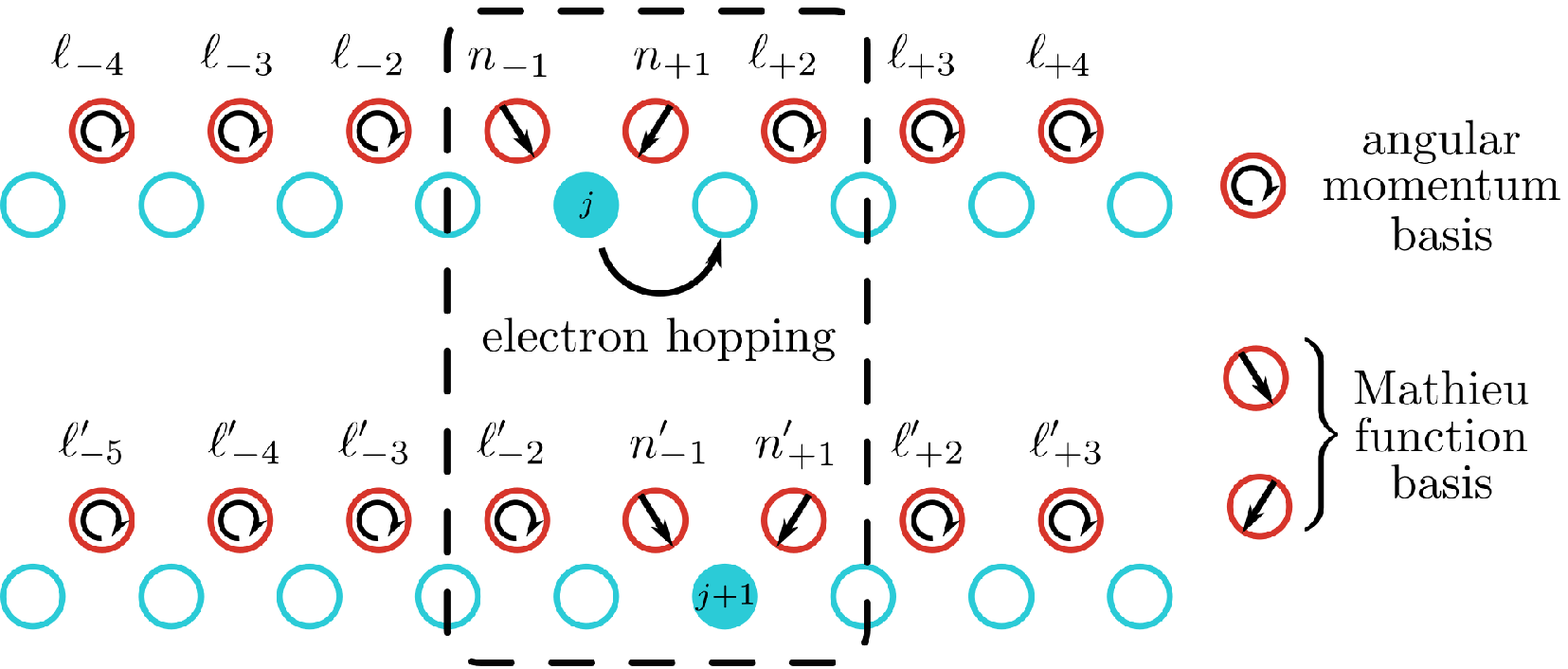} \caption{Schematic illustration of the dipole-state-dependent correlated tunneling process described within the pseudo-Lang-Firsov transformation formalism. Blue empty (filled) circles label empty (occupied) electron sites. Red circles show the dipole positions modeled by planar rotors. The arrows and notation $n_s$, $\ell_s$, show the employed basis type (see legend) and state for the particular rotor site, $s = \pm 1, \dots, \pm M/2$. The tunneling of the electron from site $j$ to  $j+1$ couples the states of the rotors within the dashed rectangle while for the remaining sites $\ell_{s-1}'=\ell_{s}$ holds.  \label{fig:Lang_Firsov}} \end{figure}

In our case, however,  \(\hat{H}_{V_0 \to \infty}\) yields the Mathieu equation \citeS{AbramowitzStegun} and therefore the operator \(\hat{U}\) has no simple analytic form. To describe how a pseudo-Lang-Firsov transformation can be performed in our case, let us assume that that the electron is localized at position $j$. This allows us to diagonalize the rotor sector of the strong-coupling Hamiltonian  by solving the corresponding Mathieu and free-rotor equations. Notice that within its eigenbasis the strong-coupling Hamiltonian reads \begin{equation} \hat{H}_{V_0 \to \infty} = \sum_{j = 1}^M \sum_{k=1}^{\infty} \epsilon_{k} |j \rangle \langle j | \otimes | \Psi^{\rm rotors}_{j,k} \rangle \langle \Psi^{\rm rotors}_{j,k} |, \end{equation} where, $\epsilon_k$, $| \Psi^{\rm rotors}_{j, k} \rangle$ are the eigenenergies and eigenstates of the rotor system respectively. Here we have used the fact that due to translational invariance $\epsilon_k$ is independent of the position of the electron.  Furthermore, the electronic and rotor wavefunctions are in a product state but $| \Psi^{\rm rotors}_{j, k} \rangle$ depends on the electron position, $j$, since only the dipoles next to the electron interact with it. The eigenstates of \(\hat{H}_{V_0 \to \infty}\) form a complete basis and thus the identity operator can be expanded as \(\hat{\mathbb{I}} = \sum_{j, k} |j \rangle \langle j | \otimes | \Psi^{\rm rotors}_{j,k} \rangle \langle \Psi^{\rm rotors}_{j,k} |\). This allows us to express the dipole--electron Hamiltonian of Eq.~(1) as \begin{equation} \begin{split} \hat{H} =& -t\sum_{j=1}^{\infty} \sum_{k,l}  \bigg( \langle \Psi^{\rm rotors}_{j,k} | \Psi^{\rm rotors}_{j+1,l} \rangle|j \rangle \langle j + 1 | \\ &\hspace{2cm}\otimes | \Psi^{\rm rotors}_{j,k} \rangle \langle \Psi^{\rm rotors}_{j + 1,l} | + {\rm h.c.} \bigg) \\ &+ \sum_{j=1}^{\infty} \sum_{k}\epsilon_{k} |j \rangle \langle j | \otimes | \Psi^{\rm rotors}_{j,k} \rangle \langle \Psi^{\rm rotors}_{j,k} |. \end{split} \label{pLFHamilt_rotors} \end{equation} By appropriately selecting the many-rotor state indices \(k\) and \(l\) and by making use of the translational invariance we can simplify Eq.~\eqref{pLFHamilt_rotors} further, see  Fig. \ref{fig:Lang_Firsov}. Here, the indices \(n^k_{\pm 1} = 0, 1, \dots\) parametrize the single particle eigenstate of the dipoles neighbouring the electron, \(| \psi^{\rm M}_{n^k_{\pm 1}} \rangle\), which solves the corresponding Mathieu equation. Furthermore, \(\ell^k_{j} = 0, \pm 1, \dots\), with \(j = \pm 2, \pm 3, \dots\), are the indices of the angular momentum eigenstates, \( \hat{L}_z | \psi^{\rm AM}_{\ell^k_{j}} \rangle = \ell^k_j | \psi^{\rm AM}_{\ell^k_{j}} \rangle \), for the dipoles further that the neighbouring ones.   Note that within this framework each different many-rotor state $k$ corresponds to a unique configuration $\{n^k_{\pm 1}, \ell^k_{\pm 2}, \ell^k_{\pm 3}, \dots \}$ of the above mentioned single-rotor states. With these definitions the overlaps of the many-rotor states contributing to tunneling read \begin{equation} \label{LFoverlaps} \begin{split} \langle \Psi^{\rm rotors}_{j,k} |\Psi^{\rm rotors}_{j + 1,l} \rangle =&  \left( \prod_{j' = -2,\pm 3, \pm 4, \dots} \delta_{\ell^k_{j'} \ell^l_{j' - 1}} \right) \\ &\times\langle \psi^{\rm M}_{n^k_{-1}} | \psi^{\rm AM}_{\ell^l_{-2}} \rangle \langle \psi^{\rm M}_{n^k_{+1}} | \psi^{\rm M}_{n^l_{-1}}  \rangle \\ &\times \langle \psi^{\rm AM}_{\ell^k_{+1}} | \psi^{\rm M}_{n^l_{+1}} \rangle \equiv O^R_{k,l}, \end{split} \end{equation} which are  independent of \(j\) and thus translationally invariant. Also, $O^{L}_{k l} = \langle \Psi^{\rm rotors}_{j + 1,k} |\Psi^{\rm rotors}_{j,l} \rangle = (O^{R}_{l k})^*$ holds. By transforming to the momentum basis for the electron, \(| q \rangle = \frac{1}{\sqrt{M}} \sum_{j=1}^M e^{i \frac{2 \pi q}{M} j} | j \rangle\), Eq.~\eqref{pLFHamilt_rotors} reduces to \(\hat{H}=\sum_{q = 0}^{M-1} \hat{H}_q \otimes |q \rangle \langle q |\), with \begin{equation} \begin{split} \hat{H}_q =& -t \sum_{k,l} \left( O^{R}_{k,l} e^{i \frac{2 \pi q}{M}} | \Psi^{\rm rotors}_k \rangle \langle \Psi^{\rm rotors}_l | + {\rm h.c.} \right) \\ &+ \sum_k \epsilon_k |\Psi^{\rm rotors}_k \rangle \langle \Psi^{\rm rotors}_k |, \end{split} \label{pLFHamilt_rotor_rotor_int} \end{equation} describing the $t$-dependent effective interactions among the rotors when the electron lies in a particular \(| q \rangle\) state. The energy of the $k$-th many-rotor state is the sum of the energies of the constituent single-rotor states, namely \begin{equation} \begin{split} \epsilon_k =& \frac{B}{4} \left[ f_{n^k_{+1}}\left(\frac{2 V_0}{B} \right) + f_{n^k_{-1}}\left(\frac{2 V_0}{B} \right) \right] \\ &+ B \sum_{j = 2}^{M/2} \left[\left(\ell^k_j\right)^2 + \left(\ell^k_{-j}\right)^2\right], \end{split} \end{equation} with \(f_n(q)\) given in terms of the Mathieu characteristic numbers, namely \(f_n(q) = a_n(q)\) for even \(n\) and \(f_n(q) = b_n(q)\) for odd \(n\).

For \(B \gg t\) and $V_0 \sim B$ it follows that \(t \ll \epsilon_k - \epsilon_{k'} \), for any $k$, $k'$, and the $t$-mediated interaction does not affect the state of the rotors to a large degree. Thus the different tunnelling channels \(| \Psi_k^{\rm rotors} \rangle\) do not interfere with one another and define  different bands of the polaron with energies \(\varepsilon_k(q) = \epsilon_k -2 t|O^R_{kk}| \cos(\frac{2 \pi q}{M} + \arg(O^R_{k,k}))\). For small $V_0 < B$, states with $n_{\pm 1}^k \neq n_{\pm 1}^{k'}$ but $\ell_{j}^k = \ell_{j}^{k'}$ for all $j = \pm 2, \pm 3, \dots$ are coupled, implying local fluctuations of the rotor state in the vicinity of the electron, leading to interference of the above mentioned tunnelling channels.  However, the number of such coupled states is independent of $M$, and the pseudo-Lang-Firsov approach can efficiently describe the polaron state.

In contrast, for \(B \ll t\), which is the relevant case for applications in perovskites, \(t \gg \epsilon_k - \epsilon_{k'} \) holds independently of $V_0$ at least in the cases where $n_{\pm 1}^k = n_{\pm 1}^{k'}$ and 
\begin{equation}
\sum_{j = 2}^M \left[ (\ell_{j}^k)^2 - (\ell_{j}^{k'})^2 \right] + \sum_{j = 2}^M \left[ (\ell_{-j}^k)^2 - (\ell_{-j}^{k'})^2 \right] \ll \frac{t}{B}.
\end{equation}
Therefore, in this case an extensive number of different \(| \Psi_k^{\rm rotors} \rangle\) states are strongly coupled by $t$-dependent effective interactions, provided that $V_0 \neq 0$ and thus $O^R_{kl} \neq \delta_{kl}$ hold. Consequently, the description of the system in this pseudo-Lang-Firsov basis becomes complicated. That is the main reason for the development of the vGH ansatz approach allowing for the construction of a Lang-Firsov-type basis that takes into account the effect of $t$-mediated interactions in a variational optimal manner.

\section{Perturbative treatment of the rotor-lattice Hamiltonian}

In order to get an insight into the polaron state let us now consider the case where $V_0$ is much smaller than the rest of the system parameters and can thus be treated perturbatively. To this end, within this section we apply the Brillouin-Wigner (BW) perturbation theory \citeS{Hubac2010}, which treatment as we will see below can be used to infer the results of other commonly used theoretical approaches in polaron physics.

Note that for $V_0 = 0$ the rotational and translational degrees of freedom decouple and as such we can define their eigenstates as $| \mathbf{m} = (m_1, m_2, ..., m_M) \rangle$ and $|k \rangle$, respectively. Here, $k$ denotes the quasimomentum of the electron. The corresponding eigenenergies are $\epsilon^{\rm rot}_{\mathbf{m}} = B \sum_{i = 1}^M m_{i}^{2}$ and $\epsilon^{\rm tr}_{k} = -2 t \cos(k)$. The interaction Hamiltonian acting here as the perturbation, see also Eq.~(1), reads
\begin{equation}
\begin{split}
    \hat{H}_I = &- V_0 \sum_{j = 1}^M \hat{a}_j^{\dagger} \hat{a}_j \cos \left(\phi_j + \frac{\pi}{4} \right) \\
         &- V_0 \sum_{j = 1}^M \hat{a}_j^{\dagger} \hat{a}_j \cos \left(\phi_{j - 1} - \frac{\pi}{4} \right).
\end{split}
\end{equation}
Given that all interaction terms appearing in the rotor-electron interaction Hamiltonian are of the form $\hat{a}_n^{\dagger} \hat{a}_n e^{\pm i \phi_j}$ this implies that only the states $| k ; \mathbf{m} \rangle$ and $| k ; \mathbf{m} \pm \hat{\mathbf{e}}_{j} \rangle$, where $\hat{\mathbf{e}}_j$ is the unit vector of the $j$-th axis, are directly coupled by the interaction. Therefore, within the second-order BW perturbation theory the wavefunction expansion reads
\begin{equation}
\begin{split}
| \Psi (k,\mathbf{m}) \rangle = \alpha_{k,\mathbf{m}} | k; & \mathbf{m}  \rangle \\
+ \sum_{\mathbf{k}'} \sum_{j = 1}^M \bigg( 
    &\beta_{k,k',\mathbf{m},j}  | k'; \mathbf{m} + \hat{\mathbf{e}}_{j} \rangle \\
+ &\gamma_{k,k',\mathbf{m},j} | k'; \mathbf{m} - \hat{\mathbf{e}}_{j} \rangle
    \bigg),
\end{split}
\label{BrillouinWignerAnsatz}
\end{equation}
where the wavefunction coefficients $a_{k, \mathbf{m}}$, $\beta_{k,k',\mathbf{m},j}$ and $\gamma_{k,k',\mathbf{m},j}$ are expressed in terms of the total energy of the system, $E$, as
\begin{equation}
\begin{split}
\alpha_{k,\mathbf{m}} &= \sqrt{Z}, \\
\beta_{k,k',\mathbf{m},j} &= - \sqrt{Z} \frac{\langle k';\mathbf{m}+\hat{\mathbf{e}}_j | \hat{H}_I | k; \mathbf{m} \rangle}{\epsilon^{\rm tr}_{k'} + \epsilon_{\mathbf{m}+\hat{\mathbf{e}}_j}^{\rm rot} - E}, \\
\gamma_{k,k',\mathbf{m},j} &= - \sqrt{Z} \frac{\langle k';\mathbf{m}-\hat{\mathbf{e}}_j | \hat{H}_I | k; \mathbf{m} \rangle}{\epsilon^{\rm tr}_{k'} + \epsilon_{\mathbf{m}-\hat{\mathbf{e}}_j}^{\rm rot} - E}.
\end{split}
\label{BWWavefunction}
\end{equation}
Note that the wavefunction renormalization of the perturbative state is performed via the insertion of the of the polaron residue, $Z$. This factor is calculated by demanding that $\langle \Psi(k, \mathbf{m}) | \Psi(k, \mathbf{m}) \rangle = 1$. The above lead to the following equation for the polaron energy
\begin{equation}
E = \epsilon^{\rm tr}_{k} + \epsilon^{\rm rot}_{\mathbf{m}} - \Sigma_{k,\mathbf{m}}(E),
\label{BWEnergyEquation}
\end{equation}
where $\Sigma_{k,\mathbf{m}}(E)$ denotes the so-called self energy of the system
\begin{equation}
\begin{split}
\Sigma_{k,\mathbf{m}}(E) = \sum_{k'} \sum_{j=1}^{M} \bigg(&\frac{ | \langle k; \mathbf{m} | \hat{H}_I | k';\mathbf{m}+ \hat{\mathbf{e}}_j \rangle |^{2} }{\epsilon^{\rm tr}_{k'} + \epsilon_{\mathbf{m}+\hat{\mathbf{e}}_j}^{\rm rot} - E} \\
+& \frac{ | \langle k; \mathbf{m} | \hat{H}_I | k';\mathbf{m}- \hat{\mathbf{e}}_j \rangle |^{2} }{\epsilon^{\rm tr}_{k'} + \epsilon_{\mathbf{m}-\hat{\mathbf{e}}_j}^{\rm rot} - E} \bigg).
\end{split}
\end{equation}

Importantly,  Eq.~\eqref{BWEnergyEquation} can also be derived by using Eq.~\eqref{BrillouinWignerAnsatz} as a variational ansatz and minimizing the energy functional $E = \langle \Psi (k,\mathbf{m}) | \hat{H} | \Psi (k,\mathbf{m}) \rangle$, under the constraint of normalized $| \Psi (k,\mathbf{m}) \rangle$. This approach is commonly referred to as the Chevy ansatz approach \citeS{Chevy2006} and has applications in Fermi-polarons emerging in ultracold atomic Fermi gases \citeS{ScazzaValtolina2017,KohstallZaccanti2012,SchirotzekWu2009,CetinaJag2016}. Since Eq.~\eqref{BWEnergyEquation} is derived within BW perturbation theory, it features, in principle, multiple solutions corresponding to the analytic continuation of each of the participating $V_0 = 0$ eigenstates, $| k ; \mathbf{m} \rangle$. In addition, since it can be derived within the Chevy ansatz, the lowest in energy solution of Eq.~\eqref{BWEnergyEquation} is an upper bound to the true ground state energy of the system, corresponding to the polaron. Within this framework we can identify several polaronic properties such as the above mentioned residue $Z$, the polaron energy $E_p = E - \epsilon^{\rm tr}_{k} = -\Sigma_{k,\mathbf{0}}(E_{p} + \epsilon^{\rm tr}_{k})$ and the polaronic effective mass $m^{*} = (\frac{\partial^2 E}{\partial k^2})^{-1}$.

To proceed note that the matrix elements of $\hat{H}_I$ read
\begin{equation}
\begin{split}
\langle k' ; \mathbf{m} + \hat{\mathbf{e}}_j | \hat{H}_I | k; \mathbf{m} \rangle &= \\
-\frac{V_0 e^{i (k - k')\left( j - \frac{1}{2} \right)}}{M} & \cos \left( \frac{k - k'}{2} + \frac{\pi}{4} \right) \\
\langle k' ; \mathbf{m} - \hat{\mathbf{e}}_j | \hat{H}_I | k; \mathbf{m} \rangle &=  \\
 -\frac{V_0 e^{i (k - k')\left( j - \frac{1}{2} \right)}}{M} & \cos \left( \frac{k - k'}{2} - \frac{\pi}{4} \right),
\end{split}
\end{equation}
and consequently, the self energy for $\mathbf{m} = \mathbf{0}$ reads
\begin{equation}
\begin{split}
\Sigma_{k,\mathbf{0}} (E) & = \frac{V_0^2}{M} \sum_{k'} \frac{1}{B -E - 2 t \cos k'} \\
& = \left\{
\begin{array}{r c l}
- \frac{V_0^2}{\sqrt{(B-E)^2-4 t^2}} & \text{for} & E > B + 2 t, \\
\frac{V_0^2}{\sqrt{(B-E)^2-4 t^2}} & \text{for} & E < B - 2 t. 
\end{array}
\right. 
\end{split}
\end{equation}
In the intermediate range $B - 2 t < E < B + 2t$ the self-energy becomes imaginary, indicating that no polaron exists in this regime. This stems from the extrapolation to the thermodynamic limit by substituting $\sum_{k} \to \frac{M}{2 \pi} \int \mathrm{d}k$. In this limit, the bands corresponding to the rotor excitations become a continuum of states in the energy interval $B - 2 t < E < B + 2t$. Thus any discrete state that couples to this continuum of excitations becomes exponentially damped in time explaining its imaginary self-energy.

Having an exact expression for $\Sigma_{k,\mathbf{m}}(E)$ we can identify the minimum of the polaron band. To find the minimum of the energy we differentiate $E$ with respect to $k$ for $E< B -2t$, yielding
\begin{equation}
\left( 1 + \frac{\partial \Sigma_{k,\mathbf{0}}}{\partial E} \right) \frac{\mathrm{d} E}{\mathrm{d} k} = \frac{\mathrm{d} \epsilon_k^{\rm tr}}{\mathrm{d} k} - \frac{\partial \Sigma_{k,\mathbf{0}}}{\partial k}.
\end{equation}
Therefore, $k =0$ is an extremal point since $\partial \Sigma_{k,\mathbf{0}}/\partial k =0$, $\mathrm{d} \epsilon_k^{\rm tr} /\mathrm{d} k = 0$ and $\partial \Sigma_{k,\mathbf{0}} /\partial E >0$. Using the above and by differentiating once more with $k$ we find
\begin{equation}
\left( 1 + \frac{\partial \Sigma_{k,\mathbf{0}}}{\partial E} \right) \frac{\mathrm{d}^2 E}{\mathrm{d} k^2} = \frac{\mathrm{d}^2 \epsilon_k^{\rm tr}}{\mathrm{d} k^2}.
\end{equation}
Thus, we conclude that $k = 0$ is the minimum of the polaron band for all values of $B$, $V_0$ and $t>0$.

\begin{figure}
    \centering
    \includegraphics[width=1.0\columnwidth]{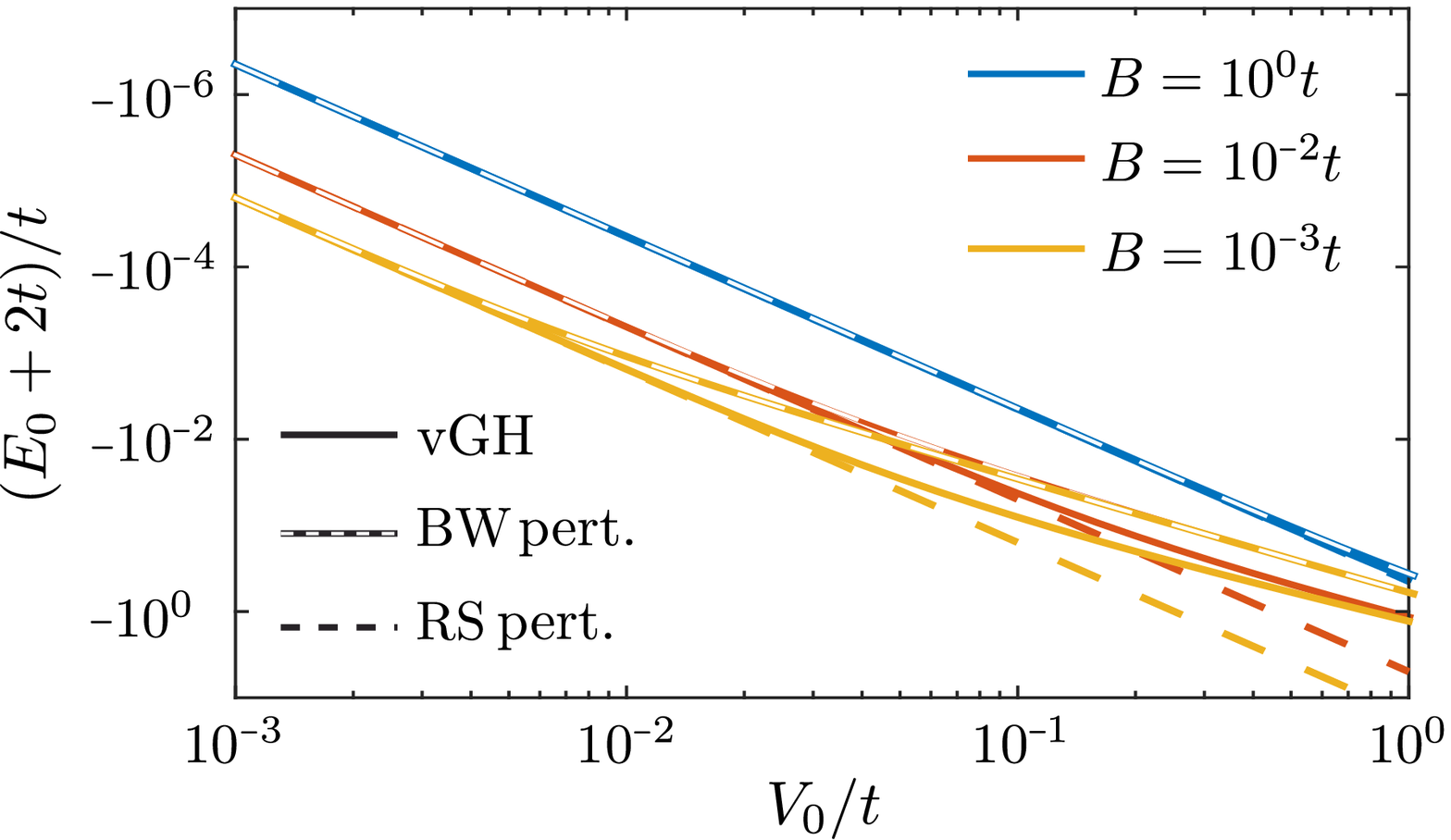}
    \caption{
(a, b) Comparison of the polaron energy, $E_0$, obtained by the vGH approach and Brillouin-Wigner and Reileigh-Schr{\" o}dinger perturbation theory for different values of $B$ as a function of $V_0$. The vGH results correspond to \(M = 1024\), while perturbation theory refers to $M \to \infty$.}
    \label{fig:energy_BW}
\end{figure}

The above allows us to evaluate the polaron characteristics by focussing on $k = 0$. First, the polaron energy is the lowest in energy solution of the algebraic equation 
\begin{equation}
E_p = - \frac{V_0^2}{\sqrt{(B+2t-E_p)^2 - 4 t^2}}.
\label{BWPolaronEnergy}
\end{equation}
Which up to fourth order in $V_0$ yields
\begin{equation}
E_p = - \frac{V_0^2}{\sqrt{B(B+4t)}} + \frac{B+2t}{B^2(B+4t)^{2}}V_0^4 + \mathcal{O}(V_0^6).
\label{BWPolaronEnergyTaylor}
\end{equation}
By substituting $E = \epsilon_k^{\rm tr}$ in the right hand size of Eq.~\eqref{BWEnergyEquation} it can be shown that the above expansion up to order $\propto V_0^2$ agrees with the second-order Reileigh-Schr{\" o}dinger perturbation theory. For this reason we employ $E_p = -V_0^2/\sqrt{B(B+4t)}$ as a proxy of the perturbative result in the main text. Note  that here  by employing Eq.~\eqref{BWPolaronEnergy} it can be proven that within the Chevy ansatz $E_p + V_0^2/\sqrt{B(B+4t)} > 0$ holds for all values of the parameters $B$, $V_0$ and $t$.  In addition, explicit numerical solutions of Eq.~\eqref{BWPolaronEnergy}, see Fig.~\ref{fig:energy_BW}, show that the vGH value of $E_p$ is always significantly smaller than the Chevy ansatz result demonstrating that the vGH approach is a significant improvement to the Chevy ansatz. 

Nevertheless, Eq.~\eqref{BWPolaronEnergyTaylor} indicates that the characteristic interaction scale obtained via BW perturbation theory is $V_0/\sqrt{B(B+4t)}$. Indeed, it can be seen that the vGH results presented in the main text begin to deviate when this dimensionless scale becomes of order $\sim 1$. Finally, let us derive the value of the effective mass within the above-mentioned perturbation theories. Within BW perturbation theory/Chevy ansatz the effective mass is a function of the polaron energy, $E_p$
\begin{equation}
\frac{m^{*}_{p}}{m^{*}} = 1 - \frac{B + 2t - E_p}{V_0^4} E_p^3.
\label{BWMass}
\end{equation}
And thus a substitution of Eq.~\eqref{BWPolaronEnergyTaylor} to Eq.~\eqref{BWMass} yields up to fourth order in $V_0$
\begin{equation}
\begin{split}
\frac{m_p^{*}}{m_p} =& 1 + \frac{B +2 t}{[B(B + 4t)]^{3/2}} V_0^2 \\
&- \frac{2 (B^2 + 4 B t + 6 t^2)}{B^3(B+4t)^3} V_{0}^4 + \mathcal{O}(V_0^6).
\end{split}
\label{BWMassTaylor}
\end{equation}
The same evaluation within the Reileigh-Schr{\" o}dinger perturbation theory results to
\begin{equation}
\begin{split}
\frac{m_p^{*}}{m_p} =& 1 + \frac{2 t}{[B(B + 4t)]^{3/2}} V_0^2 + \mathcal{O}(V_0^4).
\end{split}
\end{equation}
Therefore, the results for the effective mass agree up to quadratic order for $B \ll t$. In the main text, though, we employ $m_p^{*}/m_p = 1 + (B +2 t)V_0^2/[B(B + 4t)]^{3/2}$ as the perturbative result we compare with with vGH since it provides improved agreement even in the case of $B = t$.

\section{Details on the vGH approach}

\subsection{The vGH equations of motion}

To variationally evaluate the polaron ground-state and to get insight into the linear-response dynamics of the rotor lattice model described by Eq.~(1) we resort to the Dirac-Frenkel variational formalism, $E[\varphi_1(\phi;\tau), \dots, \varphi_M(\phi;\tau)] = \langle \Psi(\tau) | \hat{H} - \frac{\partial}{\partial \tau} | \Psi(\tau) \rangle$ \citeS{Dirac1930annihilation,frenkel1934wave}. The Dirac-Frenkel variational principle is a time-dependent variational technique, widely employed in quantum chemistry (see e.g. \citeS{BeckJackle2000}), that allows the dynamical explorations of complex systems in a variationally optimal manner in addition to their ground state properties. Note that we have chosen the Dirac-Frenkel variational principle solely based on the fact that our analysis becomes more transparent.  Indeed, it can be shown that our variationally obtained equations of motion can be obtained by the Langrangian \citeS{kramer1981geometry,Kull2000generalized} or McLachlan \citeS{McLachlan1964variational} variational principles.  This is a consequence of the fact that the Gross-Hartree ansatz of Eq.~(2) is a linear combination of Hartree products and therefore it defines an analytic function, namely a linear combination of exponentials, due to the Thouless theorem \citeS{Broeckhove1988equivalence,Thouless1960stability}.

The energy functional stemming from the Gross-Hartree ansatz reads 
\begin{equation}
\begin{split}
E[\varphi_1(\phi;\tau), &\dots, \varphi_M(\phi;\tau)] = \\ 
    -&t \bigg(
        e^{i \frac{2 \pi q}{M}} \prod_{j=1}^M \int \mathrm{d}\phi~\varphi_{j+1}^{*}(\phi;\tau) \varphi_j(\phi;\tau) \\
        +&e^{-i \frac{2 \pi q}{M}} \prod_{j=1}^M \int \mathrm{d}\phi~\varphi_j^{*}(\phi;\tau) \varphi_{j+1}(\phi;\tau)
    \bigg) \\
    -&B \sum_{j=1}^M \int \mathrm{d}\phi~\varphi^{*}_j(\phi;\tau) \frac{\partial^2}{\partial \phi^2} \varphi_j(\phi;\tau) \\
    -&i \hbar \sum_{j=1}^M \int \mathrm{d}\phi~\varphi^{*}_j(\phi;\tau) \frac{\partial}{\partial \tau} \varphi_j(\phi;\tau) \\
    +&V_{0} \bigg[ 
         \int \mathrm{d}\phi~\cos\left(\phi + \frac{\pi}{4} \right) | \varphi_1(\phi;\tau) |^2 \\
        &+\int \mathrm{d}\phi~\cos\left(\phi - \frac{\pi}{4} \right) | \varphi_M(\phi;\tau) |^2 
    \bigg]\\
    +& \sum_{j = 1}^M \lambda_j(\tau) \left( 1 - \int \mathrm{d}\phi~|\varphi_j(\phi;\tau)|^2 \right),
\end{split}
\label{vGH_functional}
\end{equation}
with \(\lambda_j\) referring to the Lagrange multipliers ensuring the normalization of \(\varphi_j(\phi;\tau)\).

The equations of motion are obtained via varying $E[\varphi_1(\phi;\tau), \dots, \varphi_M(\phi;\tau)]$ with respect to $\varphi^{*}_j(\phi;\tau)$ and read
\begin{equation}
\begin{split}
i \hbar \frac{\partial}{\partial \tau} \varphi_j(\phi;\tau) &= 
    (\hat{H}_j - \lambda_j(\tau)) \varphi_j(\phi;\tau) \\
 &-t e^{i \frac{2 \pi q}{M}} \mathcal{T}_{jL}(\tau) \varphi_{j-1}(\phi;\tau)\\
 &-t e^{-i \frac{2 \pi q}{M}} \mathcal{T}_{jR}(\tau) \varphi_{j+1}(\phi;\tau),
\end{split}
\label{vGH_eom}
\end{equation}
where $\hat{H}_j = - B \frac{\partial^2}{\partial \phi^2}  + \delta_{j1} V_0 \cos\left(\phi + \frac{\pi}{4} \right)  + \delta_{jM} V_0 \cos\left(\phi - \frac{\pi}{4} \right)$ and  the non-linearity of the above-equations stems from the mean-field tunneling couplings  $\mathcal{T}_{jL}(\tau) = \prod_{k \neq j} \int \mathrm{d}\phi~\varphi_k^{*}(\phi;\tau) \varphi_{k - 1}(\phi;\tau)$,  $\mathcal{T}_{jR}(\tau) = \prod_{k \neq j} \int \mathrm{d}\phi~\varphi_k^{*}(\phi;\tau) \varphi_{k + 1}(\phi;\tau)$ which are analogous to the \(O^R_{k,l}\) appearing in the Lang-Firsov formalism, see Eq.~\eqref{LFoverlaps}. Finally, in order to calculate the Langrange coefficients we demand that the \(\varphi_j(\phi;t)\) functions remain normalized and employ the fact that the Hamiltonian is Hermitian to obtain
\begin{equation}
\begin{split}
   \lambda_j(\tau) =& \int \mathrm{d}\phi~\varphi^{*}_j(\phi;\tau) \hat{H}_j \varphi_j(\phi;\tau) \\
   &-t \left( e^{i \frac{2 \pi q}{M}} \mathcal{T}_{L}(\tau) + e^{-i \frac{2 \pi q}{M}} \mathcal{T}_{R}(\tau) \right),
\end{split}
\end{equation}
where  $\mathcal{T}_{L}(\tau) = \prod_{k = 1}^M \int \mathrm{d}\phi~\varphi_k^{*}(\phi;\tau) \varphi_{k - 1}(\phi;\tau)$ and  $\mathcal{T}_{R}(\tau) = \prod_{k = 1}^M \int \mathrm{d}\phi~\varphi_k^{*}(\phi;\tau) \varphi_{k + 1}(\phi;\tau)$.  The above expression implies that $\lambda_j(t)$ is always real, and thus even if the $\lambda_j(\tau) \varphi_j(\phi; \tau)$ term of Eq.~\eqref{vGH_eom} is neglected, the magnitude of the single rotors states is conserved, since
\begin{equation}
    \frac{\mathrm{d}}{\mathrm{d} \tau} \left[ \int \mathrm{d}\phi \left| \varphi_j(\phi;\tau) \right|^2 \right] = \frac{2}{\hbar} \mathcal{I}(\lambda_j(\tau)) =0.
\end{equation}
Therefore the Lagrange multipliers are not {\it per se} needed for dynamical investigations, e.g. to study polaron dynamics.

However, herewith we are mainly interested in the ground state properties of the system, which can be calculated by imaginary time propagation.  Within this approach we perform the transformation $\tau \to -i \tau$ in Eq.~\eqref{vGH_eom} resulting in a diffusion equation. This equation has an important property that the energy of the propagated state monotonically decreases in time according to $\sim e^{-(E(\tau) - E_0)\tau}$, where $E_0$ is the true ground state energy, and therefore the ground state is obtained in the limit of  $\tau \to \infty$. In our implementation we perform finite imaginary time propagation up to the point that the right-hand side of Eq.~\eqref{vGH_eom} is smaller than a tolerance of $10^{-12} B$, thus ensuring that the final state is stationary with a confidence comparable to the machine error.

\subsection{Comparison with exact diagonalization for small systems}

As discussed in the main text, the vGH ansatz of Eq.~(2), neglects dipole--dipole correlations stemming from the effective rotor--rotor interactions due to electron tunnelling. We expect that these corrections are small and such they do not significantly affect the behavior of the system. Accounting for such correlations is a non-trivial task, as they involve multiple configurations of rotor states. In the absence of approximations, i.e. within exact diagonalization (ED), there are $\mathcal{M}^M$ such different configurations, where $\mathcal{M}$ is the number of single-rotor states considered. To ensure the convergence of the ED, $\mathcal{M}$ should be large enough so that the observables of interest become independent of its increase. This implies an exponential increase of the numerical complexity with the system size, and consequently the ED treatment is computationally prohibitive for large $M$. Therefore, to obtain a numerical estimate of the error in the vGH results due to neglecting these correlations, we have to rely on small systems where ED is feasible.

In particular, for our ED calculations we used $\mathcal{M} = 21$ resulting in $194481$ and $85776121$ rotor configurations for $M = 4$ and $M = 6$ respectively. The individual single-rotor states correspond to the eigenstates of the $\hat{L}_z$ operator, with eigenvalues $| \ell_z | < (\mathcal{M}-1)/2$. This choice of the many-rotor basis is sufficient for the ED energies to converge at the $10^{-5}$ level.

\begin{figure}
    \centering
    \includegraphics[width=1.0\columnwidth]{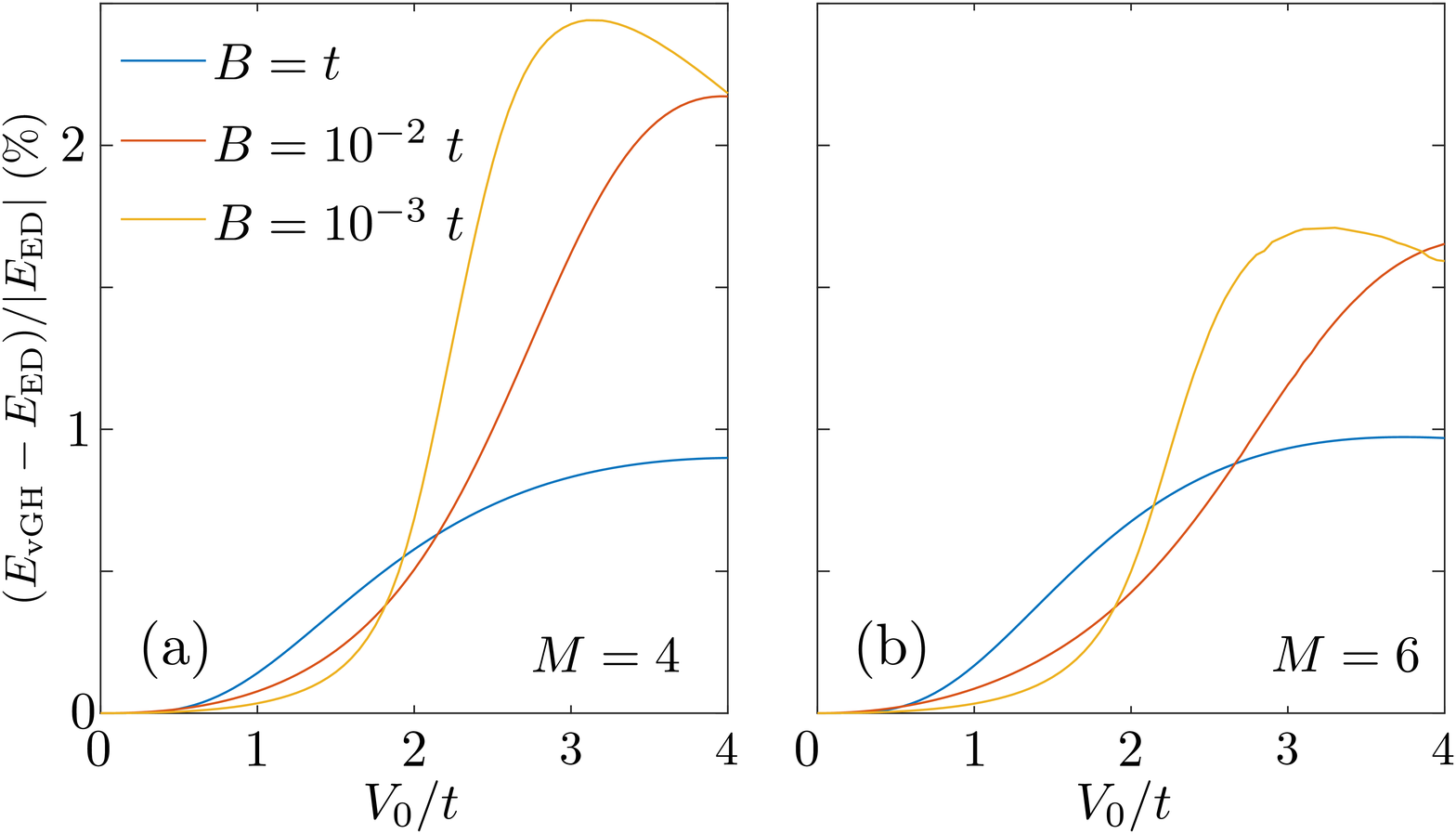}
    \caption{Percentage deviation of the ground state energy within the ED and vGH approaches, $(E_{\rm vGH} - E_{\rm ED})/|E_{\rm ED}|$, as a function of $V_0$ for (a) $M=4$ and (b) $M=6$ for different $B/t$ (see legend).}
    \label{fig:benchmark_energies}
\end{figure}

The percentile deviation of the vGH and ED ground state energies, $(E_{\rm vGH} - E_{\rm ED})/|E_{\rm ED}|$ is shown in Fig.~\ref{fig:benchmark_energies}. Here it is verified that the energy contribution of the rotor-rotor correlations is indeed small, lying in the few $\%$ range. In particular, we observe that the deviation is the largest in the interaction regime where the ferroelectric domain-wall forms, $V_0 > 2 t$. In this regime $(E_{\rm vGH} - E_{\rm ED})/|E_{\rm ED}|$ additionally exhibits an increasing tendency with decreasing $B$. In contrast, the rotor-rotor correlations seem to become less significant as $B$ decreases for interactions supporting the polarized state, $V_0 < 2 t$. Our results further suggest that the correlation corrections become less pronounced for increasing $M$, compare Fig.~\ref{fig:benchmark_energies}(a) and Fig.~\ref{fig:benchmark_energies}(b), provided that $B < 10^{-2}$.

\begin{figure}
    \centering
    \includegraphics[width=1.0\columnwidth]{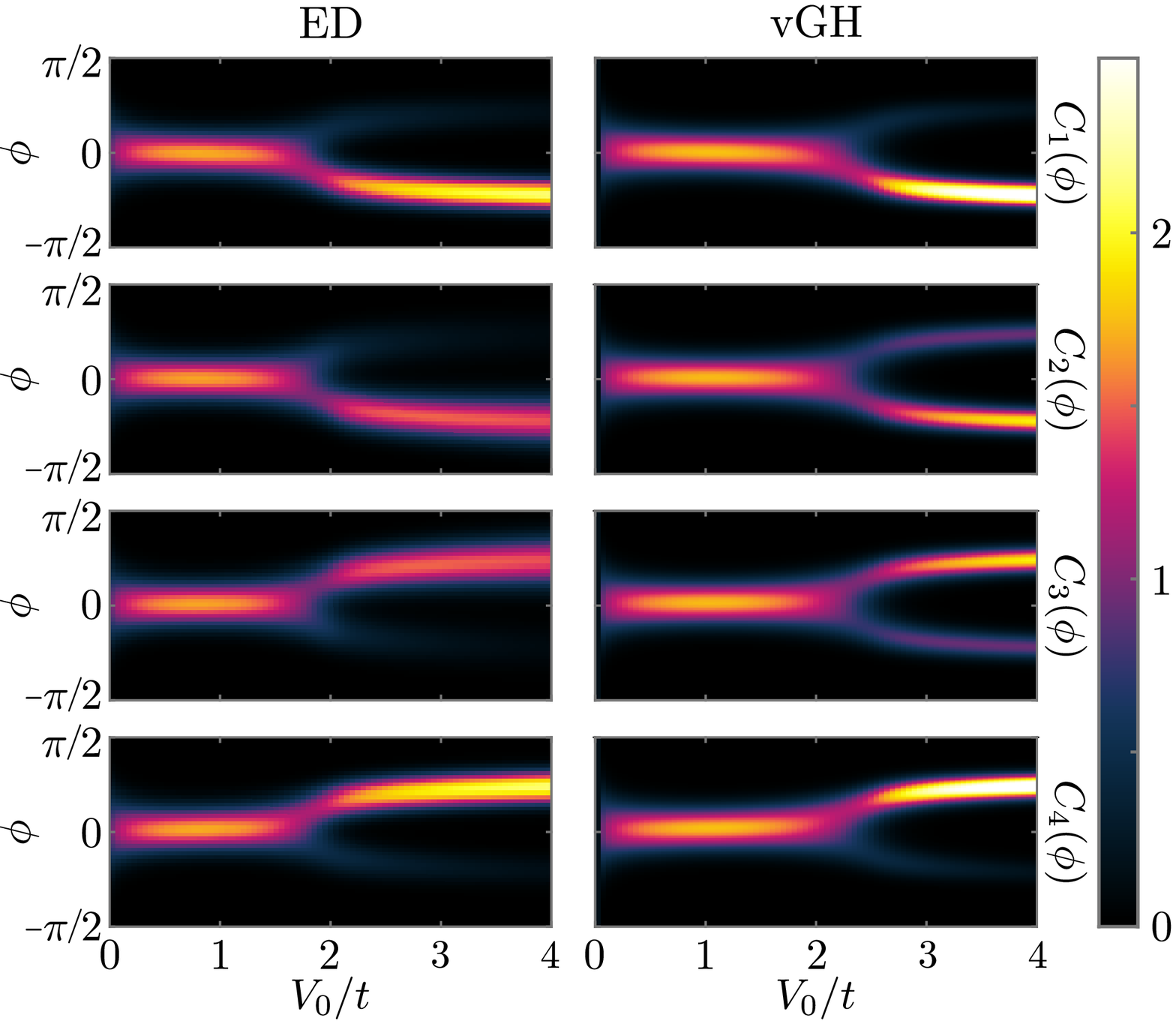}
    \caption{Rotor--electron correlation functions, $C_j(\phi)$, with $j = 1, \dots, M$ (see row labels) as a function of $V_0$ within the exact diagonalization (left column) and vGH (right column) approaches. In both cases a small system with $M=4$ and $B = 10^{-3} t$ is considered.}
    \label{fig:benchmark_densities}
\end{figure}

To demonstrate that our results are robust to the inclusion of rotor-rotor correlations Fig.~\ref{fig:benchmark_densities} compares $C_j(\phi)$ within ED and vGH. The behavior of $C_j(\phi)$ for the different approaches is nearly identical qualitatively, but there are a few notable quantitative differences.
In particular, while both approaches capture the crossover from the almost perfectly ferroelectrically polarized to the domain-wall state, the threshold shifts to a lower $V_0$ value within ED.
In addition, within the $V_0 > 2 t$ where the domain-wall forms, vGH shows significantly larger values of $C_2(\phi)$ and $C_3(\phi)$ than ED for $\phi = \pi/4$ and $\phi = -\pi/4$ respectively.
The above indicates that the overlap of adjacent rotors decreases when accounting for rotor-rotor correlations, which can be associated with a reduction of the mean-field tunneling integrals $\mathcal{T}_{j L}$ and $\mathcal{T}_{j R}$. 
Therefore, rotor--rotor correlations might induce further increase of the polaron effective mass when the domain wall forms.

In conclusion, rotor--rotor interactions do not substantially alter the polaron state, however, properly accounting for them might be beneficial for obtaining high-accuracy predictions for the polaronic properties.

\bibliographystyleS{apsrev4-1}
\bibliographyS{sup_bibliography}
\end{document}